\documentclass{emulateapj}
\usepackage{color,amsmath}
\newcommand{\fAGN}{\ensuremath{f({\rm AGN})_{\rm MIR}}\,}
\newcommand{\fTOT}{\ensuremath{f({\rm AGN})_{\rm total}}\,}

\begin{document}

\title{The role of star-formation and AGN in dust heating of {\it \MakeLowercase{z}} = 0.3-2.8 galaxies - I. Evolution with redshift and luminosity}
\author{Allison Kirkpatrick\altaffilmark{1}, Alexandra Pope\altaffilmark{1}, Anna Sajina\altaffilmark{2}, Eric Roebuck\altaffilmark{2}, Lin Yan\altaffilmark{3}, Lee Armus\altaffilmark{4},
Tanio D\'{i}az-Santos\altaffilmark{5}, Sabrina Stierwalt\altaffilmark{6}}
\altaffiltext{1}{Department of Astronomy, University of Massachusetts, Amherst, MA 01002, USA, kirkpatr@astro.umass.edu}
\altaffiltext{2}{Department of Physics \& Astronomy, Tufts University, Medford, MA 02155, USA}
\altaffiltext{3}{Infrared Processing and Analysis Center, California Institute of Technology, Pasadena, CA 91125, USA}
\altaffiltext{4}{Spitzer Science Center, California Institute of Technology, Pasadena, CA 91125, USA}
\altaffiltext{5}{N\'{u}cleo de Astronom\'{i}a de la Facultad de Ingenier\'{i}a, Universidad Diego Portales, Av. Ej\'{e}rcito Libertador 441, Santiago, Chile}
\altaffiltext{6}{Department of Astronomy, University of Virginia, Charlottesville, VA 22904, USA}
\begin{abstract}
We characterize infrared spectral energy distributions of 343 (Ultra) Luminous Infrared Galaxies from $z=0.3-2.8$. We diagnose the presence of an AGN by decomposing individual {\it Spitzer} mid-IR spectroscopy into emission from star-formation and an AGN-powered continuum; we classify sources as star-forming galaxies (SFGs), AGN, or composites. Composites comprise 30\% of our sample and are prevalent at faint and bright $S_{24}$, making them an important source of IR AGN emission. We combine spectroscopy with multiwavelength photometry, including {\it Herschel} imaging, to create three libraries of publicly available templates (2-1000\,$\mu$m). We fit the far-IR emission using a two temperature modified blackbody to measure cold and warm dust temperatures ($T_c$ and $T_w$).  We find that $T_c$ does not depend on mid-IR classification, while $T_w$ shows a notable increase as the AGN grows more luminous. We measure a quadratic relationship between mid-IR AGN emission and total AGN contribution to $L_{\rm IR}$. AGN, composites, and SFGs separate in $S_8/S_{3.6}$ and $S_{250}/S_{24}$, providing a useful diagnostic for estimating relative amounts of these sources. We estimate that $>$40\% of IR selected samples host an AGN, even at faint selection thresholds ($S_{24}>100\,\mu$Jy). Our decomposition technique and color diagnostics are relevant given upcoming observations with the James Webb Space Telescope.
\end{abstract}

\section{Introduction}
Internally, galaxy evolution is driven by ongoing star formation and an active galactic nucleus (AGN), and these two processes often
occur simultaneously in massive galaxies. Evolved galaxies formed most of their stellar and black hole mass in the era $z\sim1-3$, making high redshift sources quintessential for disentangling how the growth of an AGN impacts the interstellar medium (ISM) of an actively star forming galaxy \citep[and references therein]{madau2014}.
The majority of the black hole activity in the early Universe is occurring behind dust screens, as evidenced by the largely unresolved cosmic X-ray background at energies $>6$\,keV \citep{hickox2007}. In addition, the bulk of the star formation during this period is occurring in Luminous ($L_{\rm IR}>10^{11}{\rm L}_{\odot}$) and Ultra Luminous Infrared Galaxies ($L_{\rm{IR}}>10^{12}{\rm L}_{\odot}$), known as (U)LIRGs \citep[e.g.,][]{murphy2011a}. (U)LIRGs at high redshift form stars at prodigious rates (star formation rate (SFR)$\ge10 - 100\,M_\odot\,{\rm yr}^{-1}$),
and many show signs of concurrent AGN growth \citep[e.g.,][]{sajina2007,pope2008,coppin2010,kirkpatrick2012}, providing an attractive option for studying the simultaneous assembly of black hole and stellar mass.

Since a large fraction of star formation and AGN activity in the early Universe is obscured by dust, it is necessary to turn to the infrared spectrum to study these processes. Ubiquitous infrared data from space telescopes has made it possible to identify star forming and AGN signatures in the dust emission. In the near-IR, H- emission from the older stellar population is visible as a stellar bump, peaking at 1.6\,$\mu$m. However, if an AGN is present, it can heat the surrounding torus to $T\gtrsim1000\,$K, causing the dust to radiate into the near-IR and obscure the stellar bump \citep{donley2012}.
The mid-IR spectrum is the most rich for identifying AGN and star formation signatures, as it contains dust and gas emission/absorption lines and an underlying continuum. 
The most prominent dust emission complexes are produced by polycyclic aromatic hydrocarbons (PAH); PAHs are abundant in galaxies with metallicity close to solar, such as high redshift (U)LIRGs \citep{magdis2012} . PAHs are excited by UV and optical photons and are primarily located in star forming regions; as such, PAHs are good tracers of the rate of star formation in a galaxy \citep{peeters2004}. Additionally, the mid-IR spectrum may exhibit continuum emission coming from very small dust grains stochastically heated by the interstellar radiation field, or a stronger, steeply rising continuum due to emission from a hot dusty torus enveloping the AGN.

The bulk of IR luminosity is emitted in the far-IR, which is comprised of  warm dust (T$\sim60-100$\,K) and cold dust (T$\sim$20\,K) components; the temperatures and relative amounts of each component are an excellent indicator of the dominant power source in a galaxy \citep{kirkpatrick2012}. The warm and cold dust components arise from different locations in the ISM \citep{dunne2001}. The cold dust is located in the diffuse ISM and is emission from large dust grains and the bulk of the dust mass. The warm dust emanates in star forming regions, or is possibly heated by radiation from an AGN.

Even with the availability of an abundance of infrared data, observations at high redshift are still limited due to either confusion limits of telescopes or long required integration times for faint
galaxies. As a result, a common technique is to apply local templates to scant photometry for distant galaxies in order to extrapolate information about their star formation rates, $L_{\rm IR}$, or dust masses. In particular, many authors scale the appropriate \citet{chary2001} template to a 24\,$\mu$m photometric point to estimate $L_{\rm IR}$. However, this technique has been shown to overestimate $L_{\rm IR}$ at $z>1.5$, likely due to the changing nature of ULIRGs \citep[e.g.,][]{nordon2010,magnelli2011,elbaz2011}.
The \citet{chary2001} templates were derived from local galaxies, and local ULIRGs are almost exclusively undergoing a major merger.
A major merger of two galaxies triggers a spatially compact burst of star formation and the subsequent growth of the AGN \citep{sanders1996}.
Out to $z\approx1.5$, the SFR and specific star formation rate (sSFR=SFR/$M_\ast$) increases in disk galaxies as they approach neighbors, and after an interaction, luminous AGN signatures are detectable in the 
mid-infrared \citep{hwang2011,zam2011}. In contrast, at $z\sim2$, a significant fraction of ULIRGs have a disk morphology and lack any merger signatures, likely because the increased gas fractions can sustain the high SFRs without requiring a merger \citep{elbaz2011,kartaltepe2012}.

Assigning low redshift templates to high redshift sources correctly presents a serious problem for high redshift studies, and astronomers have addressed this problem by creating empirical high redshift templates from stacked spectral energy distributions (SEDs) \citep{elbaz2011,kirkpatrick2012,sajina2012,lee2013}.
In particular, \citet{kirkpatrick2012} and \citet{sajina2012} use mid-IR spectroscopy to diagnose the presence of an AGN, and then stack photometry and spectroscopy to measure the average IR emission properties of AGN and star forming galaxies (SFGs). Using mid-IR spectroscopy to identify AGN allows the authors to find AGN that might be missed at other wavelengths due to dust obscuration, and the resulting templates can then be used to assess the presence of an AGN in high redshift galaxies that have only a few photometric observations \citep{kirkpatrick2013,nelson2014,stanley2015}. \citet{kirkpatrick2012} and \citet{sajina2012} are limited by the number of galaxies with available mid-IR spectroscopy. Due to sample size, those studies mainly compare the IR properties of SFGs and AGN, but do not focus on composite sources which have a mixture of both star formation and AGN activity. 

In this paper, we extend the work of \citet{kirkpatrick2012} and \citet{sajina2012} by combining the individual samples to create a large sample of 343 high redshift (U)LIRGs. With this combined sample, we are able to probe the effect of a growing AGN on the observed SED by classifying sources as SFGs, AGN, and composites. We are also able to quantify changes in the dust emission as a function of redshift and $L_{\rm IR}$. We create three libraries of empirical IR SED templates, which we make publicly available. Our sample is {\it unique} in that all of our sources have mid-IR spectroscopy, allowing us to robustly separate AGN from SFGs. With our statistically significant sample and template libraries, we investigate how the dust properties, such as temperatures and heating sources, vary as the AGN grows more luminous. In Section \ref{sec:two}, we describe our sample and data sets, and in Section \ref{sec:three}, we discuss our mid-IR decomposition technique, which allows us to determine the presence/strength of an AGN. In Section \ref{sec:four}, we present our three empirical template libraries. In Section \ref{sec:five}, we discuss the relationship between AGN signatures in the mid-IR and the total contribution of an AGN to $L_{\rm IR}$. In Section \ref{sec:six}, we consider how the dust properties of high redshift (U)LIRGs, as indicated by our templates, relate to AGN growth and galaxy evolution. Finally, we summarize our findings in Section \ref{sec:seven}. Throughout this paper, we adopt a standard cosmology with $H_{0}=70\,\rm{km}\,\rm{s}^{-1}\,\rm{Mpc}^{-1}$, $\Omega_{\rm{M}}=0.3$, and $\Omega_{\Lambda}=0.7$. 

\section{Sample and Observations}
\label{sec:two}
\subsection{Sample Description}
We have assembled a multi-wavelength data set for a sample of 343 high redshift ($z\sim0.3-2.8$) (U)LIRGs in the Great Observatories Origins Deep Survey North 
(GOODS-N),
Extended {\it Chandra} Deep Field Survey (ECDFS), and {\it Spitzer} Extragalactic First Look Survey (xFLS) fields. All sources are selected to have mid-IR spectroscopy from the {\it Spitzer Space Telescope} Infrared Spectrograph (IRS), 
necessary to concretely quantify the IR AGN emission in each galaxy. Our sample contains a range of sources from individual observing programs, each with differing selection criteria. However, the overarching selection criterion is that each galaxy must be bright enough at 24\,$\mu$m (observed frame) to be detectable in ($<10$ hours). More specific properties of the different fields are outlined below.

The xFLS sample is comprised of archival sources with IRS spectroscopy \citep[complete sample details can be found in][]{sajina2012}. 
The sources were selected to have an observed 24\,$\mu$m flux density greater than 0.9\,mJy and to have an R magnitude of $m_{R,{\rm Vega}}\geq 20$. {\it Spitzer} Program IDs and references are listed in Table \ref{tbl:pid}. 
The xFLS IRS sample contains just under half of the xFLS sources that meet the above photometric criteria; however, \citet{sajina2012} find that the IRS sample has the same $S_{24}/S_{8}$ color distribution as the parent sample and is representative of a 24\,$\mu$m-selected sample ($>$\,0.9\,mJy) at $z$\,$\gtrsim$1. The $m_{R,{\rm Vega}}>20$ criteria removes the $z$\,$\sim$\,0.2 peak found in the redshift distribution of a purely 24\,$\mu$m-selected sample.

The GOODS-N and ECDFS samples include all sources in these fields that were observed with {\it Spitzer} IRS \citep[complete details are in][]{kirkpatrick2012}. All of these sources were selected at 24\,$\mu$m (observed), and 93\% of have $S_{24}>100\,\mu$Jy. The IRS sources occupy the same regions in $S_{250}/S_{24}$ and $S_8/S_{3.6}$ colorspace as the parent MIPS 24\,$\mu$m GOODS sample with $S_{24}>100\,\mu$Jy, and they have a similar redshift distribution as those MIPS sources in GOODS for which we have redshift estimates ($\sim750$ sources).

\begin{figure}
\centering
\includegraphics[width=3.3in]{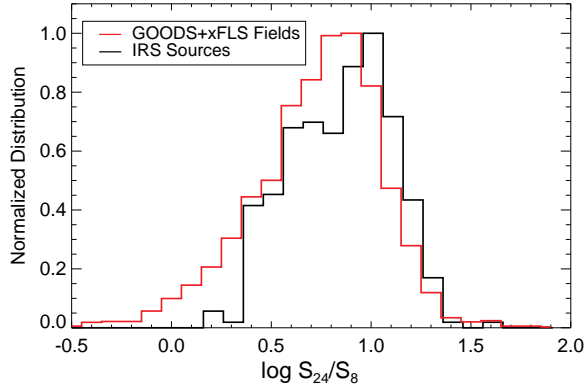}
\caption{The distribution in $S_{24}/S_8$ (observed) for our IRS sample (black) compared with the full xFLS and GOODS fields (red). We have down-weighted the distribution of xFLS sources (each source is assigned a weight of 0.03) to account for the difference in the sizes of the GOODS and xFLS fields. The IRS sample is representative of the full fields except for very blue sources ($\log S_{24}/S_8<0.1$) which are likely undetected in the far-IR. \label{fig:rep}}
\end{figure}

We illustrate the representativeness of the combined sample in Figure \ref{fig:rep}. We plot the distribution of $S_{24}/S_8$ for our IRS sources, and we compare with the full distribution of GOODS-N, GOODS-S, and xFLS sources. We are limited by the choice of color due to the different wavelength coverage and depths of the xFLS and GOODS fields. We combine $S_8$ with $S_{24}$ as this color traces the relative amount of PAH emission or silicate absorption compared with warm continuum emission, both of which we use to diagnose the presence of an AGN. The xFLS field (2.7 deg$^2$) is much larger than the GOODS fields (0.09 deg$^2$), so we have weighted the distribution of the xFLS sources by the ratio of the field areas. The distributions are not consistent, which is a natural result of our combining several samples with different selection criteria. However, our IRS sample is generally representative of the GOODS and xFLS fields in the $S_{24}/S_8$ color, although there is a subset of sources with low $S_{24}/S_8$ ratios that we are missing.
It is important for the reader to bear in mind that for this study, we are interested in sources that have mid-IR spectroscopy and PACS or SPIRE photometry. Sources with low $S_{24}/S_8$ ratios are likely to be very faint at longer wavelengths. 
This sample is representative of sources that are detected both in the mid-IR and far-IR and may not cover the parameter space of sources fainter than our flux limits in either IR regime. 

An additional result of our different $S_{24}$ selection criteria for fields of different sizes is that we have biased our redshift distribution of SFGs and AGN. Our selection of SFGs is predisposed towards strong PAH emitters at redshifts $z\sim1$ and $z\sim2$, where the PAH features fall in the 24\,$\mu$m bandpass. Strong AGN are intrinsically brighter at $S_{24}$ \citep{kirkpatrick2012}; due to the smaller area of the GOODS fields, our brightest AGN are found in the xFLS field at $z>2$. However, owing to the bright detection limits, no SFGs are found at similar redshifts in xFLS. We use the IRS spectrum to determine redshifts (Section \ref{sec:red}), and this introduces a bias as well, since we require coverage of PAH features or the 9.7\,$\mu$m silicate absorption feature. In our sample, 60\% of sources have coverage of the 6-8\,$\mu$m PAH complexes, 64\% have coverage of the 11.2-12.7\,$\mu$m complexes, and 82\% have coverage of the 9.7\,$\mu$m silicate absorption feature. We have 36 sources with a featureless spectrum that have optically available redshifts, but we have rejected a further
$\sim10\%$ of sources that meet our selection criteria because they have featureless spectra and no reliable optical redshift.

\begin{deluxetable*}{lcl}
\tablecolumns{3}
\tablecaption{IRS Sample \label{tbl:pid}}
\tablehead{\colhead{PID} & \colhead{\# of Sources\tablenotemark{a}} & \colhead{References}}
\startdata
20629	&	136								& \citet{dasyra2009}\\
30431 &	49+15 \,									& \citet{fadda2010} \\
3748	&	39					& \citet{yan2007}\\
20456 &	22+2 \,									&	\citet{pope2008,murphy2009} \\
40918 &		13+2 \,								&	\nodata \\
20733   &    10+2 \,									&	\nodata \\
288	&	6 									&	\citet{pope2013} \\
3216 &	6 									&	\nodata \\
20083	& 6		& \citet{lacy2007} \\
15 & 5 & \citet{weedman2006,martinez2008} \\
252 	&	2+2									&	\citet{teplitz2007} \\
20081 &	4									&	\citet{menendez2009} \\
30419 &	4									&	\citet{donley2010} \\
20128	& 4	& \nodata \\
50419	&	2								&	\nodata \\
50305	 &	3								&	\nodata \\
30447	& 2	& \nodata \\
3223	 &	1									& \citet{sturm2006} \\
531 & 1	&  \citet{carilli2010,riechers2014} \\
20542	& 1	& \nodata \\
20767 &	1									&	\nodata \\
50324	&	1								&	\nodata \\	
50512	&	 1								&	\nodata \\
50647	&	1								&	\nodata
\enddata
\tablenotetext{}{We list the {\it Spitzer} Program ID (PID) of our sources, 
the number of sources from that program, and a reference when available. We stress that we have reduced all data from 
these programs ourselves and apply our own spectral decomposition in a consistent manner.}
\tablenotetext{a}{A few programs resulted in so-called bonus sources which were not a part of the initial target list. We list the number of bonus sources after the ``+" sign.}

\end{deluxetable*}

\subsection{Spectroscopy and Photometry}
Full details on the IRS observations and data reduction of the xFLS sources are discussed in \citet{dasyra2009}. Here we only present a brief summary. The data reduction starts with the {\it Spitzer} Basic Calibrated Data (BCD). We removed the residual median sky background from each IRS low-resolution order (the short-low (SL) order covering 5.2-14.7\,$\mu$m, and the long-low (LL) order covering  14.3-35.0\,$\mu$m). We did a mixture of automatic and manual bad pixel removal, replacing their values with interpolations from their neighbors.  The 1D spectra for each nod position and each spectral order were extracted using the {\it Spitzer} Science Center package SPICE, adopting the ``optimal" extraction technique which in essence is a weighted PSF-fitting and is recommended for faint sources. Aperture and slit-loss corrections are applied. Finally, the two nod positions are averaged and the different orders merged using linear interpolation in the overlap region. The flux calibration was found to be consistent between the orders and consistent with the broadband IRAC 8\,$\mu$m and MIPS 24\,$\mu$m flux densities.

The low resolution ($R=\lambda/\Delta\lambda\sim100$) {\it Spitzer} IRS spectra in the GOODS-N and ECDFS fields were reduced following the method detailed in \citet{pope2008}. Specifically, since many of these are long integrations,
we take care to remove latent build-up on the arrays over time,
and we create a ÒsuperskyÓ from all the off-nod observations
to remove the sky background. One-dimensional spectra are
extracted using SPICE in optimal
extraction mode. For each target, a sky spectrum is also extracted
to represent the uncertainty in the final target spectrum. The target spectrum flux calibration was found to be consistent with the broadband MIPS 24\,$\mu$m flux densities.

The xFLS field was observed with {\it Herschel} SPIRE as part of the HerMES survey, while the GOODS-N and ECDFS fields were imaged with {\it Herschel} PACS and SPIRE as part of the GOODS-{\it Herschel} Open Time Key Program. All {\it Herschel} photometric flux densities are extracted using the MIPS 24\,$\mu$m prior positions. For sources that are blended with another galaxy based on 24\,$\mu$m prior positions, we deblend by fitting two Gaussians. If a source is blended with two or more other galaxies, we reject the photometry at this wavelength. We also reject sources that result in a $\leq 1\sigma$ detection.
For the xFLS sources, we have rejected the 250\,$\mu$m photometry for 23 sources, the 350\,$\mu$m photometry for 36 sources, and the 500\,$\mu$m photometry for 46 sources due to being too blended or too faint. In the GOODS-N and ECDFS fields, we reject 26 sources at all SPIRE wavelengths for being too blended. The sources rejected span the full redshift distribution.

\begin{figure}
\centering
\includegraphics[width=3.3in]{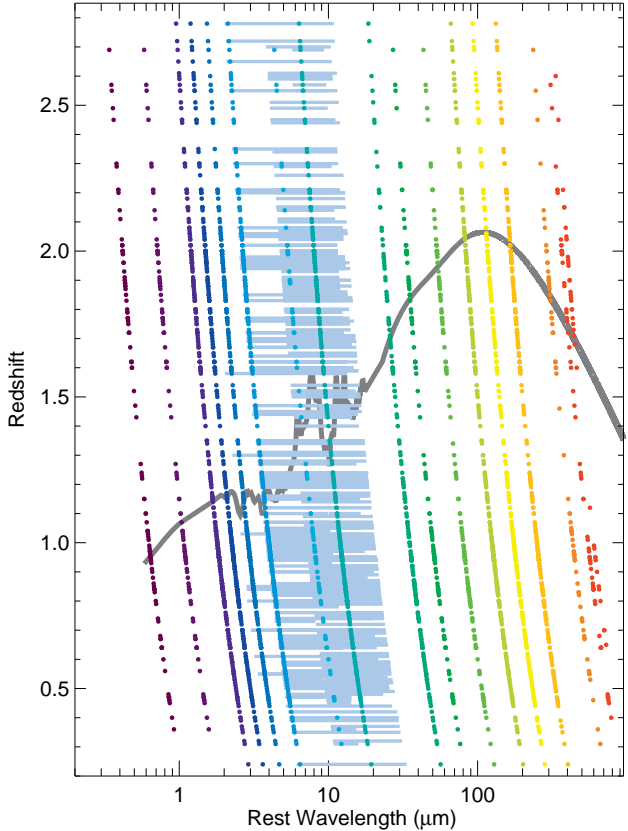}
\caption{We show the available photometry and spectroscopy for each source in our sample. We redshift the observed photometric wavelengths for individual sources to the rest frame. We plot a filled circle if a source has a photometric detection at a given wavelength, and we indicate the rest frame coverage of the IRS spectra with a blue shaded region. 
We show an IR SED in grey to better illustrate the coverage of our photometry and spectroscopy. Our spectroscopic and photometric coverage is exceptional, and there are no significant gaps in any particular bandpass due to increasing redshift. \label{fig:coverage}}
\end{figure}

We combine {\it Herschel} and {\it Spitzer} photometry and spectroscopy with ground-based near-IR and submillimeter imaging to obtain excellent coverage of the full IR spectrum from $z=0.3-2.8$. Specifically, for the GOODS-N and ECDFS sources, we have $J$ and $K$ band photometry from VLT/ISAAC \citep{retzlaff2010} and CFHT/WIRCAM \citep{wang2010,lin2012}; {\it Spitzer} IRAC 3.6, 4.5, 5.8, 8.0\,$\mu$m, IRS 16\,$\mu$m, and MIPS 24, 70\,$\mu$m imaging; {\it Herschel} PACS 100, 160 and SPIRE 250, 350, 500\,$\mu$m imaging; and 870\,$\mu$m photometry from LABOCA on APEX \citep{weiss2009} and the combined AzTEC+MAMBO 1.1mm map of GOODS-N \citep{penner2011}. For the xFLS sources, we have {\it Spitzer} IRAC 3.6, 4.5, 5.8, 8.0\,$\mu$m, and MIPS 24, 70, 160\,$\mu$m imaging; {\it Herschel} SPIRE 250, 350, 500\,$\mu$m imaging; and MAMBO 1.2\,mm imaging \citep{lutz2005,sajina2008,martinez2009}. We illustrate the wavelength coverage of our data in Figure \ref{fig:coverage}.

\section{Mid-IR Spectral Decomposition}
\label{sec:three}
\subsection{AGN Strength}
\label{sec:decomp}
We perform spectral decomposition of the mid-IR spectrum ($\sim5-18\,\mu$m rest frame) for each source in order to disentangle the AGN and star forming components.  \citet{pope2008} explain the technique in detail, and we summarize here. We fit the individual spectra with a model comprised of four
components: (1) the star formation component is represented by the mid-IR spectrum of the prototypical starburst M~82 (we verified the choice of template by comparing with the low redshift starburst template from \citet{brandl2006}, which produced the same results); (2) the AGN component is determined by fitting a pure power-law with the slope and 
normalization as free parameters; (3,4) extinction curves from the \citet{draine2003} dust models for Milky Way (MW) type dust is applied to the AGN component and star forming component. The full model is then
\begin{equation}
\label{eq:decomp}
S_\nu=N_{\rm AGN}\lambda^\alpha\,e^{-\tau_{\rm AGN}}+N_{\rm SF} S_\nu ({\rm M82})e^{-\tau_{\rm SF}}
\end{equation}
We fit for $N_{\rm AGN}$, $N_{\rm SF}$, $\alpha$, $\tau_{\rm AGN}$, $\tau_{\rm SF}$, and redshift simultaneously.

The extinction curve is not monotonic in wavelength and contains silicate absorption features, the most notable for our wavelength range being at 9.7\,$\mu$m. It is important to note that the assumption of MW dust has a non-negligible effect on the normalization of the AGN component, and a lower metallicity dust could lower the overall contribution of an AGN to $L_{\rm IR}$ \citep{snyder2013}. 
The M~82 template already contains 
some intrinsic extinction. We allow additional extinction to the SF component beyond that 
inherent in the template and find this to be necessary for 24\% of the sources. 

\begin{figure}
\centering
\includegraphics[width=3.37in]{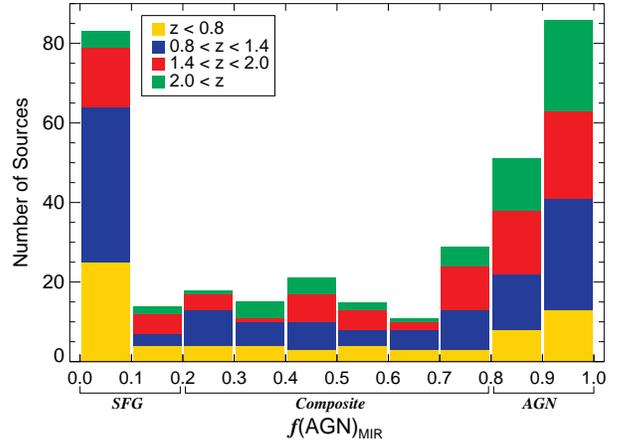}
\caption{The distribution of mid-IR AGN fraction, determined from the mid-IR spectral decomposition.
The colors correspond to redshift. A large portion (30\%) are SFGs with little AGN contribution, but there is also a sizeable
population of AGN (36\%). We indicate our \fAGN classifications below the x-axis.\label{fig:frac_dist}}
\end{figure}

For each source, we quantify the strength of the AGN, \fAGN, as the fraction of the total mid-IR
luminosity coming from the extincted power-law continuum component. We classify the sources as SFGs (\fAGN$<$0.2), composites (\fAGN=0.2-0.8), and AGN (\fAGN$>$0.8).
Figure \ref{fig:frac_dist} illustrates the \fAGN distribution of the sample, with colors corresponding to redshift. There are roughly equal numbers of SFGs (30\%), composites (34\%), and AGN (36\%). This high percentage of AGN is a selection effect due to the different field sizes and flux limits. Throughout the paper, we refer to these mid-IR spectroscopically identified AGN simply as AGN, though the reader should bear in mind that they may not be identified as such at other wavelengths.

Assessing the reliability of our decomposition technique is of utmost importance for interpreting the results in this paper. We have tested the soundness of our \fAGN in three ways: 
1) The most serious concern is between dust extinction and AGN fraction. We find that if we remove the extinction component, 70\% of our sample would have $\Delta\fAGN<0.1$, while 21\% would lie within $\Delta\fAGN<0.2$. In general, not including the extinction component scatters to lower \fAGN. 2) We create synthetic spectra, where we know the input AGN fraction, and add noise. We then run our decomposition code on our synthetic spectra. We can recover \fAGN within 0.1 even at a signal to noise ratio of three. 3) We test our results by comparing to another decomposition method, deblendIRS, presented in \citet{caballero2015}. The deblendIRS technique decomposes IRS spectra into a stellar, PAH, and AGN component using a library of 19 stellar, 56 PAH, and 39 empirical AGN templates. This allows for variation in the PAH features. When comparing the two techniques, we find on average $\Delta(\fAGN-f({\rm AGN})_{\rm deblendIRS})=0$ with a standard deviation of 0.15. These three techniques underscore the reliability of the \fAGN values presented here.

\subsubsection{Comparison of AGN Indicators}
We briefly address how our AGN quantification technique compares with two other AGN selection methods often used at high redshift. Our GOODS-N and ECDFS sources have {\it Chandra} 2Ms and 4Ms (respectively) X-ray observations \citep{alexander2003,luo2008,xue2011}. Of our AGN in these fields, 73\% are detected in the X-ray. We estimate that our AGN all have comparable intrinsic X-ray luminosities, indicating that those AGN that are not detected might be Compton thick \citep{alexander2008,bauer2010}. Of our composite sources, 35\% have an X-ray detection. Eleven of our AGN are included in a study by \citet{brightman2014} that measures column density for sources in GOODS-S. Eight of these AGN have column densities of $N_H\approx10^{22}-10^{23}\,{\rm cm}^{-2}$, but the remaining 3 have $N_H>10^{24}\,{\rm cm}^{-2}$, indicating Compton thickness. Much more limited X-ray data exists for the xFLS field. Specifically, \citet{bauer2010} target 20 AGN sources with {\it Chandra} 150 ks observations. Only two sources are detected, and the remaining sources are estimated to be Compton thick. Overall, there is broad agreement between our mid-IR spectral AGN indicators and X-ray AGN indicators, although we stress that our technique will not be biased again obscured AGN which are much more prevalent at high redshift \citep[e.g.,][]{treister2010}.

{\it Spitzer} IRAC color selection is also commonly used to cull AGN from a sample \citep{lacy2004,stern2005,donley2012}. The criteria in \citet{donley2012} is based on colors ($S_8/S_{4.5}$ and $S_{5.8}/S_{3.6}$) that distinguish whether a galaxy has power-law emission in the near- to mid-IR, and this power-law emission is indicative of an AGN. However, in \citet{kirkpatrick2013}, we demonstrated that AGN residing in high redshift (U)LIRGs do not universally display power-law emission in these colors due to contamination from the host galaxy. 75\% of the AGN in this sample have colors indicative of an AGN according to \citet{donley2012}, while only 29\% of composites meet these criteria. The benefit of our mid-IR spectral decomposition is that we can identify heavily obscured AGN and {\it quantify} the strength of the AGN emission.

\subsection{Spectroscopic Redshifts}
\label{sec:red}
We determine redshifts for the majority of our sample by fitting the positions of the main PAH features (6.2, 7.7, 11.2, 12.7\,$\mu$m complexes). Out of our sample, 36 sources have a featureless mid-IR spectrum. In these cases, we adopt available optical spectroscopic redshifts for the GOODS/ECDFS sources \citep[e.g.,][]{szokoly2004,barger2008,popesso2009,stern2012}. Optical redshifts for the xFLS sources were determined with targeted Keck and Gemini follow-up observations \citep[e.g.,][]{choi2006,yan2007,sajina2008}. Redshifts derived from fitting the PAH features have typical
uncertainties of $\Delta z = 0.01-0.03$ \citep{dasyra2009} while redshifts based only on the 9.7\,$\mu$m silicate feature (as is the case for many of our strong AGN) 
have uncertainties of $\Delta z =0.1-0.2$ \citep{sajina2007}.

\begin{figure}
\centering
\includegraphics[width=3.4in]{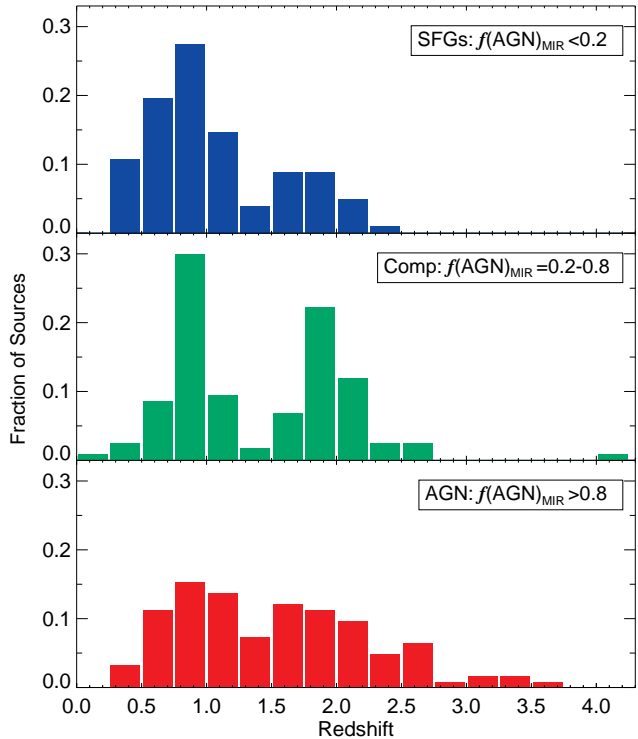}
\caption{The redshift distribution of our sample where we have separated sources by \fAGN. The top panel shows the fraction of SFGs in each redshift bin, the middle panel shows the fraction of composites, and the bottom panel shows the fraction of AGN per bin. The highest redshift sources are mainly AGN, 
which reflects the 24\,$\mu$m selection criterion, since AGN are
typically brighter at this wavelength than SFGs. The bimodal distribution which peaks at $z\sim1$ and $z\sim2$, particularly evident for the SFGs and composites, is also a byproduct of the 24\,$\mu$m selection, since at these redshifts
prominent PAH features fall in the 24\,$\mu$m bandpass. \label{fig:agn_red}}
\end{figure}

The redshift distribution is illustrated in Figure \ref{fig:agn_red}, where we separate sources according to \fAGN. 
The redshift distribution is largely bimodal, with peaks around $z\sim1$ and
$z\sim2$, which reflects the overarching 24\,$\mu$m selection criterion. At $z=1,2$, prominent PAH features fall within the 24\,$\mu$m bandpass causing an increase of detected sources with intense star formation. Conversely, at $z\sim1.5$, the 9.7\,$\mu$m silicate absorption feature falls within the 24\,$\mu$m bandpass, resulting in a dearth of sources. The highest redshift sources ($z>2.5$) are predominantly AGN; this is also a byproduct of the 24\,$\mu$m selection criterion since AGN activity boosts mid-IR emission. We have relatively more composites at $z\sim2$ than SFGs because the composites tend to be more luminous at 24\,$\mu$m due to AGN emission and are more easily detected.

\section{A New Public Library of Empirical Infrared Templates}
\label{sec:four}
The spectral energy distributions of dusty high redshift ULIRGs are seen to differ from the SEDs of local ULIRGs \citep[e.g.,][]{pope2006,elbaz2011,sajina2012,kirkpatrick2012}. In light of this, a library of templates designed
specifically for high redshift galaxies is required. Our large spectroscopic sample and wealth of multiwavelength data is ideally suited for this purpose. However, our individual mid-IR spectra are noisy, and many of our sources lack complete coverage of the peak of the SED emission in the far-IR, due to confusion limits from {\it Herschel}. Therefore, we can better study the dust emission at high redshift by considering the average SED. We combine our sources to create three libraries of {\it publicly available}\footnote{http://www.astro.umass.edu/$\sim$pope/Kirkpatrick2015} empirical SED templates: 
\begin{enumerate}
\item MIR-based Library. This is a user-friendly library suited for sources with mid-IR spectroscopy.
\item Color-based Library. This is a user-friendly library ideal for sources with only IR photometry.
\item Comprehensive Library. This library best represents the intrinsic properties (\fAGN,$L_{\rm IR}$) of our sources.
\end{enumerate} 

Within each template library, we divide our sources into subsamples using criteria outlined in Sections \ref{sec:mir_temp}-\ref{sec:com_temp}.
Table \ref{tbl:properties} describes the basic properties of the subsamples comprising each template.
We begin by shifting all spectra and photometry to the rest frame. Within each subsample, we determine the median mid-IR luminosity ($5-15\,\mu$m)
and scale the individual rest frame SEDs using this value. We choose to normalize by the mid-IR luminosity because it minimizes the scatter in $L_\nu$ between galaxies at all IR wavelengths while preserving the intrinsic average luminosity of each subsample. 
  
After normalization, we average the IR data by determining the median $L_\nu$ and wavelength in differential bin sizes,
chosen so that each bin is well-populated ($>$~5 data points). In the near-IR and far-IR, where data is scarcer, we calculate rolling medians, and we treat photometric data points and spectroscopic data points the same.
For each subsample, we randomly draw sources with replacement and recalculate the normalized median 1000 times; the uncertainty on the template is then the standard deviation around the median. Because we normalize in the mid-IR,
the resulting templates exhibit little scatter in and around these wavelengths. 

We fit a two temperature modified blackbody (2T MBB) to the bootstrapped far-IR data ($>20\,\mu$m) and uncertainties in order to characterize the shape 
of the far-IR in terms of physical parameters. The 2T MBB has the form 
\begin{equation}
\label{eq:BB}
S_\nu=a_w\times \nu^{\beta}\times B_\nu(T_{w})+a_c\times \nu^{\beta}\times B_\nu(T_{c})
\end{equation}
where $B_\nu$ is the Planck function, and $T_w$ and $T_c$ are the temperatures of the warm and cold dust components, respectively. We keep the emissivity fixed at $\beta=1.5$, assuming optically thin dust. 
The choice of model is non-trivial, and we discuss alternate far-IR models in Appendix \ref{app:alternate}.
We fit for the normalization factors, $a_w$ and $a_c$, and the temperatures, $T_w$ and $T_c$, simultaneously using a $\chi^2$ minimization technique. The error bars in this regime reflect the uncertainty of the fitted parameters, including both the intrinsic scatter among sources and the photometric uncertainties in the data.
We then verify the 2T MBB fit by overplotting photometry from $850-1100\,\mu$m, observed frame. The submillimeter data is not included in the fit because it is not available for the 
majority of sources and would therefore bias the derived cold dust temperature. For all templates, the available submillimeter data agrees with the template within the photometric 
uncertainties. Our fitting technique is illustrated for one subsample in Figure \ref{fig:sed_fit}, and we show all subsamples and corresponding templates in Appendix \ref{app:temp}.
Table \ref{tbl:temp_stats} lists $T_c$, $T_w$, and $L_{\rm IR}$ of each template.

Other popular models for fitting the full IR SED include a power-law combined with a MBB \citep[e.g.,][]{casey2012} and a hot torus model combined with a 2T MBB \citep[e.g.,][]{sajina2012}. We opt not to use these models because we do not include near- and mid-IR data in our fits as this portion of each template is created through stacking the data. We tested what effect these different models have on measuring $L_{\rm IR}$ and $T_c$ and find no significant change in these parameters.

\begin{figure}
\includegraphics[scale=0.345]{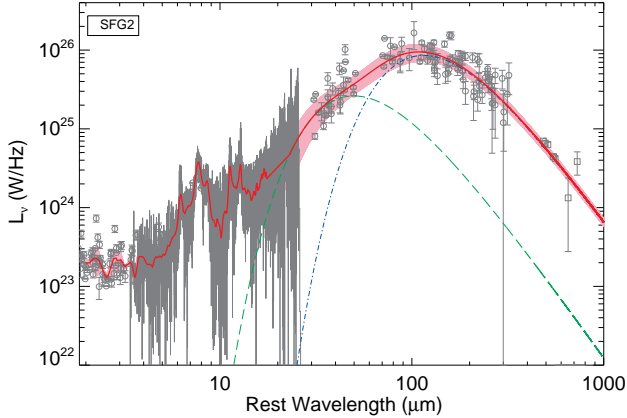}
\caption{An example of our template creation technique for the SFG2 subsample (all subsamples are listed in Table \ref{tbl:properties}). We show photometric and spectroscopic data 
(grey points and lines), and we plot the template and corresponding uncertainty in red and pink. All normalized spectroscopic and photometric data were averaged together in differential bin sizes using a bootstrapping technique to estimate the uncertainties. We then fit the far-IR averaged photometry with a 2T MBB (the green dashed line and the blue dot-dashed line represent the warm and cold dust components, respectively). We overplot available submillimeter data ($\lambda>300\,\mu$m; grey squares), {\it not included in the fit}, to check the validity of our 2T MBB fit. All data sets and corresponding templates are shown in Appendix \ref{app:temp}. \label{fig:sed_fit}}
\end{figure}

\begin{deluxetable}{lc ccc cc}
\tablecolumns{4}
\tablecaption{Categories of Template SEDs \label{tbl:properties}}
\tablehead{\colhead{Name} & \colhead{Number of Sources} & \colhead{Median $z$} & \colhead{Median \fAGN}}
\startdata
\cutinhead{MIR-based Templates (Figure \ref{fig:mir_temp})}
MIR0.0		& 68		& 0.94	& 0.00	 \\
MIR0.1		& 24		& 0.94	& 0.09		 \\
MIR0.2		& 21		& 1.10	& 0.20 \\
MIR0.3		& 16		& 1.38	& 0.28  \\
MIR0.4		& 18		& 1.39	& 0.38 \\
MIR0.5		& 21		& 0.96	& 0.49	\\
MIR0.6		& 15		& 1.59	& 0.60	 \\
MIR0.7		& 23		& 1.50	& 0.70	\\
MIR0.8		& 31		& 1.52	& 0.80 \\
MIR0.9		& 51		& 1.80	& 0.90\\
MIR1.0		& 54		& 1.18	& 1.00 \\
\cutinhead{Color-based Templates (Figure \ref{fig:col_temp})}
COLOR1	& 75		& 1.10	& 0.14 	 \\
COLOR2	& 57		& 0.94	& 0.23	 \\
COLOR3	& 41		& 1.17	& 0.39 	\\
COLOR4	& 26		& 0.95	& 0.77  \\
COLOR5	& 29		& 1.52	& 0.81 \\
COLOR6	& 25		& 1.09	& 0.96 	 \\
COLOR7	& 24		& 1.97	& 0.87 	 \\
COLOR8	& 23		& 1.83	& 0.94 \\
\cutinhead{Comprehensive Templates (Figure \ref{fig:lza_temp})}
SFG1 		& 38		& 0.92	& 0.00 	\\
SFG2 		& 23		& 0.91	& 0.00 		\\
SFG3 		& 24		& 1.75	& 0.07 		\\
Composite1 	& 24		& 0.85	& 0.38	\\
Composite2 	& 27		& 0.94	& 0.60 		\\
Composite3 	& 18		& 1.89	& 0.43	\\
Composite4 	& 29		& 1.96	& 0.57	\\
AGN1		& 22		& 0.80	& 1.00 	\\
AGN2 		& 23		& 1.03	& 0.93		\\
AGN3 		& 21		& 1.65	& 0.94 \\
AGN4 		& 31		& 1.95	& 0.93 		
\enddata
\end{deluxetable}

\begin{deluxetable}{lc ccc}
\tablecolumns{5}
\tablecaption{Properties of Template SEDs \label{tbl:temp_stats}}
\tablehead{\colhead{Name} & \colhead{$T_c$\tablenotemark{a}} & \colhead{$T_w$\tablenotemark{a}} & \colhead{$L_{\rm IR}$\tablenotemark{b}} & \colhead{$L_{\rm IR}^{\rm SF}$\tablenotemark{c}} \\
                    \colhead{} & \colhead{(K)} & \colhead{(K)} & \colhead{($10^{12}\,L_\odot$)} & \colhead{($10^{12}\,L_\odot$)}}
\startdata
\cutinhead{MIR-based Templates}
MIR0.0		& 25.7 $\pm$ 0.6	& 66.0 $\pm$ 2.4	& 0.57 $\pm$ 0.07	& 0.57 $\pm$ 0.07	\\
MIR0.1	& 26.8 $\pm$ 1.0	& 66.7 $\pm$ 4.5	& 0.72 $\pm$ 0.16	& 0.69 $\pm$ 0.15	\\
MIR0.2	& 24.6 $\pm$ 1.3	& 62.4 $\pm$ 1.4	& 1.05 $\pm$ 0.17	& 0.98 $\pm$ 0.16	\\
MIR0.3	& 27.3 $\pm$ 1.9	&\, 75.0 $\pm$ 11.3	& 1.22 $\pm$ 0.52	& 1.11 $\pm$ 0.47	\\
MIR0.4	& 29.4 $\pm$ 1.6	& 70.3 $\pm$ 3.7	& 2.21 $\pm$ 0.49 	& 1.88 $\pm$ 0.42	\\
MIR0.5	& 29.4 $\pm$ 1.8	& 84.3 $\pm$ 5.6	& 1.17 $\pm$ 0.30	& 0.92 $\pm$ 0.23	\\
MIR0.6	& 35.2 $\pm$ 3.2	& 87.7 $\pm$ 9.9	& 3.76 $\pm$ 1.43 	& 2.82 $\pm$ 1.07	\\
MIR0.7	& 26.1 $\pm$ 2.2	& 80.2 $\pm$ 3.4	& 1.95 $\pm$ 0.47 	& 1.38 $\pm$ 0.34	\\
MIR0.8	& 28.3 $\pm$ 1.3	& 85.6 $\pm$ 3.8	& 2.97 $\pm$ 0.55 	& 1.81 $\pm$ 0.34 	\\
MIR0.9 	& 29.0 $\pm$ 1.9	& 89.8 $\pm$ 6.1	& 3.27 $\pm$ 0.71 & 1.67 $\pm$ 0.36 \\
MIR1.0	& 26.3 $\pm$ 2.3	& 83.4 $\pm$ 4.5	& 1.68 $\pm$ 0.33 	& 0.72 $\pm$ 0.14	\\
\cutinhead{Color-based Templates}
COLOR1		& 26.4 $\pm$ 0.9	& 63.0 $\pm$ 4.2	& 1.16 $\pm$ 0.25 	& 1.14 $\pm$ 0.24	\\
COLOR2		& 24.8 $\pm$ 1.1	& 61.5 $\pm$ 3.4	& 0.66 $\pm$ 0.13 	& 0.59 $\pm$ 0.12	\\
COLOR3		& 26.9 $\pm$ 1.5	& 62.8 $\pm$ 4.7	& 1.89 $\pm$ 0.49 	& 1.72 $\pm$ 0.45	\\
COLOR4		& 20.9 $\pm$ 1.6	& 74.3 $\pm$ 7.4	& 0.81 $\pm$ 0.23 	& 0.52 $\pm$ 0.15	\\
COLOR5		& 28.5 $\pm$ 2.4	& 80.5 $\pm$ 4.6	& 3.35 $\pm$ 0.85 	& 2.04 $\pm$ 0.52	\\
COLOR6		& 27.0 $\pm$ 2.4	& 87.3 $\pm$ 4.6	& 1.62 $\pm$ 0.36 	& 0.66 $\pm$ 0.15	\\
COLOR7		& 37.0 $\pm$ 3.3	& 88.3 $\pm$ 7.7	& 4.82 $\pm$ 1.66 	& 2.75 $\pm$ 0.95	\\
COLOR8		& 24.4 $\pm$ 2.4	& 88.9 $\pm$ 4.1	& 2.46 $\pm$ 0.56	& 0.81 $\pm$ 0.19	\\
\cutinhead{Comprehensive Templates}
SFG1 		& 26.3 $\pm$ 1.0	& 62.4 $\pm$ 5.9	& 0.40 $\pm$ 0.11	& 0.38 $\pm$ 0.10	\\
SFG2 		& 28.1 $\pm$ 1.3	& 64.9 $\pm$ 5.6	& 1.31 $\pm$ 0.35	& 1.27 $\pm$ 0.34	\\
SFG3 		& 26.8 $\pm$ 1.8	& 58.1 $\pm$ 6.9	& 1.35 $\pm$ 0.51	& 1.28 $\pm$ 0.49	\\
Composite1  & 25.7 $\pm$ 0.9	& 81.0 $\pm$ 5.0	& 0.49 $\pm$ 0.09	& 0.45 $\pm$ 0.08	\\
Composite2 	& 30.9 $\pm$ 2.7	& 84.3 $\pm$ 4.6 & 1.31 $\pm$ 0.51	& 1.05 $\pm$ 0.41	\\
Composite3 	& 31.1 $\pm$ 2.8	& 72.5 $\pm$ 9.6	& 1.60 $\pm$ 0.72	& 1.02 $\pm$ 0.46	\\
Composite4 	& 38.9 $\pm$ 2.9	& \, 82.8 $\pm$ 15.6	& 6.96 $\pm$ 3.34	& 5.01 $\pm$ 2.40	\\
AGN1		& 21.7 $\pm$ 2.2	& 72.7 $\pm$ 7.6 & 0.47 $\pm$ 0.17	& 0.21 $\pm$ 0.08	\\
AGN2 		& 25.3 $\pm$ 2.9	& 86.0 $\pm$ 4.4 	& 2.03 $\pm$ 0.58	& 1.24 $\pm$ 0.35	\\
AGN3 		& 31.8 $\pm$ 4.1	& 78.5 $\pm$ 9.8 	& 2.38 $\pm$ 1.18	& 0.90 $\pm$ 0.45	\\
AGN4 		& 33.4 $\pm$ 5.3	& 75.2 $\pm$ 5.8 	& 6.57 $\pm$ 2.10	& 3.22 $\pm$ 1.03	
\enddata
\tablenotetext{a}{See Equation \ref{eq:BB}.}
\tablenotetext{b}{Calculated by integrating each template from 8-1000\,$\mu$m.}
\tablenotetext{c}{Fraction of $L_{\rm IR}$ attributable to star formation. Calculated total AGN contribution to $L_{\rm IR}$, \fTOT, and scaled $L_{\rm IR}$ correspondingly to obtain $L_{\rm IR}^{\rm SF}$. See Section \ref{sec:five}.}
\end{deluxetable}

\subsection{Mid-IR Based Templates}
\label{sec:mir_temp}
\begin{figure}
\centering
\includegraphics[width=3.4in]{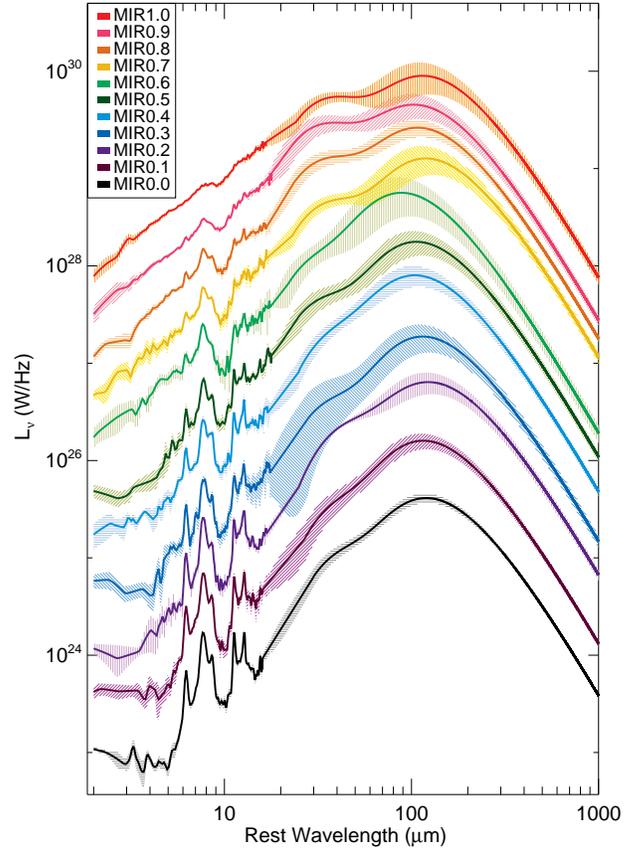}
\caption{MIR-based Template Library created by grouping sources according to \fAGN to explore how the shape of the IR SED changes as an AGN grows more luminous. Template subsample properties are listed in Table \ref{tbl:properties}. The templates have been arbitrarily offset in $L_{\nu}$ to allow for easier comparison. Shaded regions show the uncertainties for each template. MIR0.3 has particularly large uncertainties around 20\,$\mu$m, but this is due to a lack of data points in this regime. As the mid-IR AGN grows stronger, the far-IR emission becomes flatter due to an increase in the warm dust emission. \label{fig:mir_temp}}
\end{figure}

Our sample is unique in that we have mid-IR spectroscopy for every source, allowing us to classify a large sample of galaxies in a similar manner. Therefore, we
create a library of eleven templates by separating sources according to \fAGN, in order to assess what effect a mid-IR luminous AGN has on the full IR SED.
Each subsample is chosen so that it contains at least 15 sources and so that the median \fAGN increases by $\sim$0.1, spanning the range \fAGN=0.0-1.0. We list the subsample properties in Table \ref{tbl:properties} 
and show the library of MIR-based Templates in Figure \ref{fig:mir_temp}. These user-friendly templates are ideal for inferring far-IR dust properties when little or no far-IR information is available. In particular, 
this template library will be useful to derive $L_{\rm IR}$ and estimate SFRs when mid-IR spectroscopy from the forthcoming MIRI instrument on the James Webb Space Telescope becomes available.

In Figure \ref{fig:mir_temp}, we have ordered the templates by the median \fAGN of the sources that comprise each template. PAH features are visible in all but the MIR1.0 template. The MIR0.8-MIR1.0 templates all exhibit silicate absorption, although this may be a selection effect since some pure power-law spectra were excluded from the final xFLS sample.
In general, the MIR0.3-MIR0.7 subsamples contain fewer 
sources each, and these sources show a variety of SED features which is reflected in the templates and resulting errors. 
The lack of uniformity in the MIR0.3-MIR0.7 templates signals that AGN emission may manifest itself in the full IR SED differently based on some property of the host galaxy, such as the spatial distribution of the dust. In contrast, the MIR0.0-MIR0.2 templates have very small uncertainties, suggesting a remarkable uniformity in shape among SFGs. 

The MIR0.0-MIR0.2 templates are consistent in shape with the $z\sim1$ SF SED and $z\sim2$ SF SED from \citet{kirkpatrick2012}. The Silicate AGN SED from \citet{kirkpatrick2012}, created from sources with \fAGN$>$0.5 that exhibited silicate absorption at 9.7\,$\mu$m, is consistent with the MIR0.6 template. In contrast, the Featureless AGN SED from that work, created from sources with \fAGN$>$0.5 with a power-law spectrum, is not consistent with any of the templates presented here. The MIR0.8-MIR1.0 templates all have more cold dust emission than we observed previously. By combining the GOODS+ECDFS sources from \citet{kirkpatrick2012} with the xFLS sources from \citet{sajina2012}, we more than doubled the number of AGN in the sample, increasing the range of observed far-IR SEDs. We also now have proportionally more AGN with silicate absorption, rather than pure power-law AGN, and these silicate AGN tend to have more cold dust.

\begin{figure}
\centering
\includegraphics[width=3.4in]{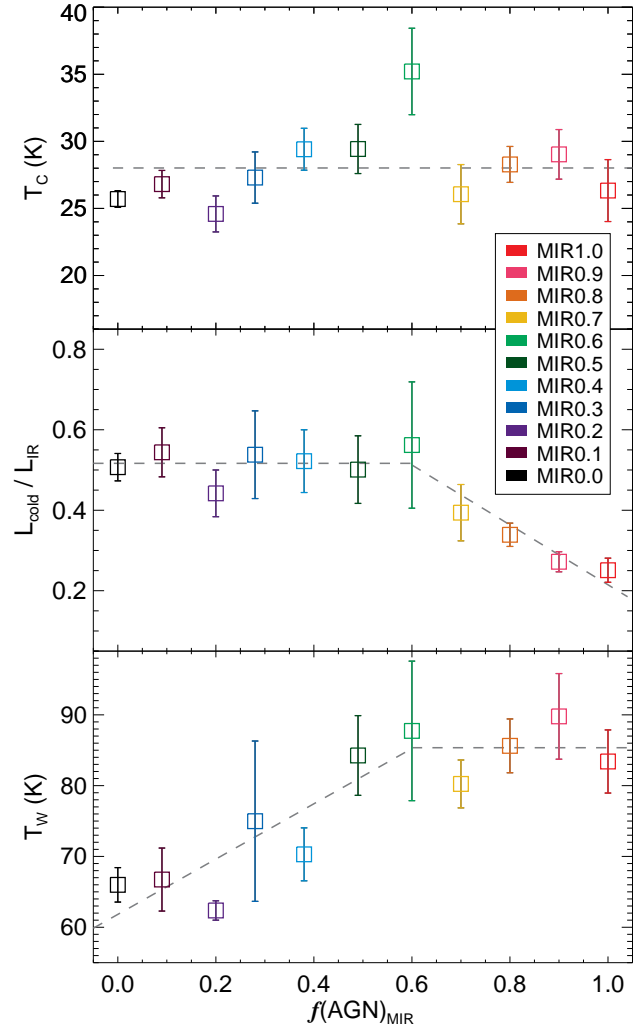}
\caption{Cold dust temperature ({\it top panel}), $L_{\rm cold}/L_{\rm IR}$ ({\it middle panel}), and warm dust temperature ({\it bottom panel}) as a function of \fAGN for the MIR-based Templates. \fAGN is the median value of the sources comprising each template. $T_w$ increases until \fAGN=0.6, while $L_{\rm cold}/L_{\rm IR}$ decreases after this point. In contrast, $T_c$ is roughly constant (dashed line is median $T_c$). \label{fig:mir_stats}}
\end{figure}

We characterize the shape of the far-IR using $T_c$, $T_w$, and $L_{\rm cold}/L_{\rm IR}$ and plot these properties as a function of \fAGN (median of each subsample) in Figure \ref{fig:mir_stats}. $L_{\rm cold}$ is derived by integrating under the cold dust MBB from Equation \ref{eq:BB}, and it arises from the diffuse ISM, making $L_{\rm cold}$ and $T_c$ secure tracers of the host galaxy \citep{dunne2001}. $T_c$ varies by less than 5\,K for almost all templates (grey dashed line is median $T_c$), illustrating that $T_c$, which quantifies the peak wavelength of the dust emission, is not correlated with the presence of a mid-IR luminous AGN. Since $T_c$ arises from the diffuse ISM, this indicates that on average, the galaxies in our sample all display extended dust emission. $T_c$ for MIR0.6 (light green) is a notable exception. $T_c$ is nearly 10\,K higher for this template, shifting the peak of the SED from $\sim110\,\mu$m to $\sim90\,\mu$m. $T_c$ is higher for MIR0.6 due to a combination of the fact that there are fewer sources in this bin, and these are the most luminous sources on average in the sample. It is possible this subsample is made up of more compact galaxies, leading to warmer overall dust temperatures. We explore correlations between $T_c$ and $L_{\rm IR}$ in Section \ref{sec:com_temp}. 

$L_{\rm cold}/L_{\rm IR}$, the fraction of $L_{\rm IR}$ due to cold dust emission, is nearly constant for the MIR0.0-MIR0.6 templates, after which it starts to decrease (middle panel of Figure \ref{fig:mir_stats}). We illustrate this trend with the grey dashed line, where we join the median $L_{\rm cold}/L_{\rm IR}$ for MIR0.0-MIR0.6 with a simple linear fit to the MIR0.6-MIR1.0 points. Until \fAGN=0.6, emission from the extended host galaxy is dominating the infrared luminosity, despite a growing contribution from an AGN to the mid-IR.

In contrast, $T_w$ increases until \fAGN=0.6, and then it is fairly constant for \fAGN=0.7-1.0 (bottom panel; dashed line is a linear fit joined to a median). $T_w$ has two possible heating sources. The first is star forming regions, either in the extended disk or in a compact starburst, although locally compact starbursts are measured to produce warmer temperatures \citep[e.g.,][]{diaz2011}.
In the MIR0.0-MIR0.1 templates, $T_w$ can be safely attributed to star formation. As the AGN grows stronger, it will contribute to $T_w$, eventually outshining any dust heated by star formation. The gas that fuels a growing AGN can fuel a compact starburst too, making it difficult to distinguish exactly what is responsible for high $T_w$ temperatures. However, the clear trend between $T_w$ and \fAGN in our sample indicates that either the AGN progressively increases its heating contribution to the wavelength range $\lambda=20-80\,\mu$m, producing higher $T_w$s, or the growth of the AGN is directly linked with a compact starburst which is responsible for the boost in $T_w$.
\fAGN=0.6 marks a turning point in the shape of the IR SED. It is here that $T_w$ reaches its peak temperature, and afterwards AGN-heated dust contributes more to $L_{\rm IR}$ than the diffuse dust heated by star formation.

The warm dust component fits to the wavelength range $\sim20-80\,\mu$m which, for our sample, is covered by MIPS and PACS observations. The xFLS sources lack PACS detections, which could affect the reliability of the warm dust fits and the trend between $T_w$ and \fAGN. We test how reliable the trend is by fitting the 2T MBB to the far-IR data after removing all PACS and MIPS 160 (available for a few xFLS sources) photometry.
The same trends between $T_w$, $L_{\rm cold}/L_{\rm IR}$ and \fAGN are observed.

\subsection{Color-Based Templates}
\label{sec:col_temp}
\begin{figure}
\centering
\includegraphics[width=3.3in]{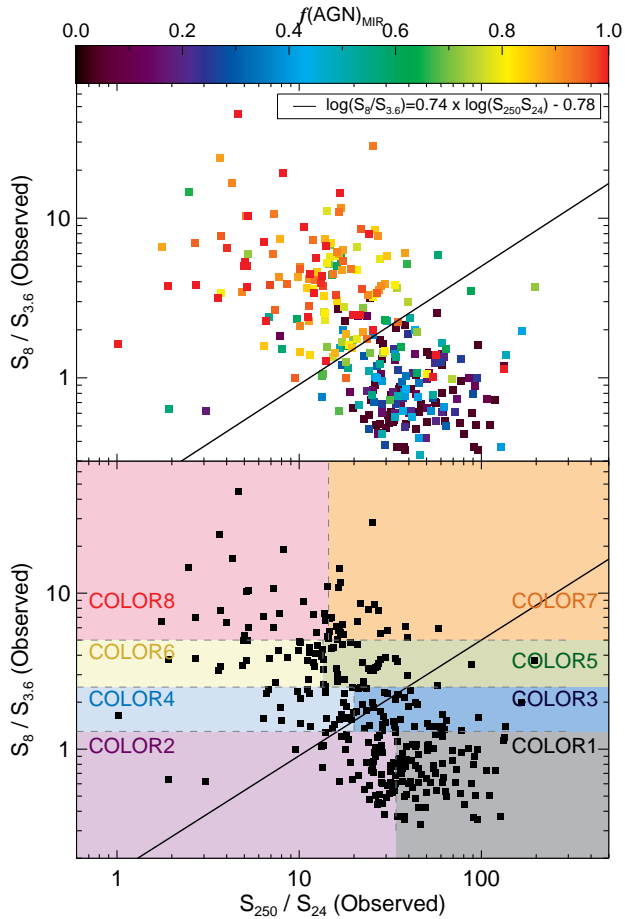}
\caption{Distribution of our sources in IR colorspace using the observed frame colors. {\it Top panel-} Each source is shaded according to \fAGN. AGN strength increases as $S_{250}/S_{24}$ decreases and $S_{8}/S_{3.6}$ increases. The dark line (equation in upper right corner) shows the empirical separation between the AGN and SFGs defined in \citet{kirkpatrick2013}. {\it Bottom panel-} For the Color-based Templates, we group sources by their location in colorspace in order to better explore the differences in the intrinsic IR SED shapes. The sources that comprise each template are illustrated by the shaded regions. \label{fig:colors}}
\end{figure}

In the MIR-based Template Library, we grouped sources according to \fAGN, but as we noted, the individual sources comprising some of the templates showed a broad range of observed SED properties.
We now explore an alternative way to sort sources and create templates based only on the SED shape of each source. In \citet{kirkpatrick2013} we created an IR 
color diagnostic designed to capture the full shape of the SED by combining far-, mid-, and near-IR photometry. We present this color diagnostic in Figure \ref{fig:colors}, where we make use of photometry from {\it Herschel} SPIRE and {\it Spitzer} MIPS/IRAC, available for 87\% of our sample.
$S_{250}/S_{24}$(observed) traces the ratio of far-IR emission to mid-IR emission, and this ratio is lower in AGN sources as the heating from the AGN boosts the mid-IR emission. 
At the redshifts of our sources, $S_{8}/S_{3.6}$(observed) is primarily tracing the stellar bump, and in this regime, radiation from the AGN washes this feature out, 
producing power-law emission. 

The top panel of Figure \ref{fig:colors} illustrates that \fAGN grows larger with decreasing $S_{250}/S_{24}$ 
and increasing $S_8/S_{3.6}$. 
There is a degree of scatter, particularly among the AGN sources, and in \citet{kirkpatrick2013} we demonstrated that much of this scatter is attributable to the broad redshift range of our sources. However, intrinsic SED shape can also produce scatter, and we have tested this effect using the library of torus models in \citet{siebenmorgen2015}. These models account for the
intrinsic
luminosity of the AGN, the viewing angle, the inner
radius of the torus, the volume filling factor and optical depth of the toroidal clouds, and the optical depth of the disk midplane in the host galaxy. We redshift the models to $z=1.5$ and plot their observed frame colors. We find that this library of AGN SEDs occupies the same general region as our AGN, although with a much broader distribution of colors, and varying the radius of the inner torus and the optical depth of the host disk does the best job at reproducing the observed scatter of our sources. Modeling the geometry of the torus in individual sources is beyond the scope of this work; however, the above suggests that allowing for a range of host galaxy optical depths can already account for much of the scatter in colorspace. Indeed, Roebuck et al.\,(2016, in prep.) uses simulations to show that our empirical IR AGN templates include not only the torus, but also the host dust-reprocessed light. We conclude that both redshift and intrinsic SED shape can account for the scatter of our sample.

To create the Color-based Templates, we divide sources according to $S_{250}/S_{24}$ and $S_8/S_{3.6}$,  
so that we can quantify differences in $T_w$ and $T_c$ as a smooth function of $S_{250}/S_{24}$ and $S_8/S_{3.6}$. We illustrate the color criteria for each subsample in the bottom panel of Figure \ref{fig:colors}.   We have blindly chosen the color criteria rather than basing them on existing knowledge of the IR SED so that we can more fairly test how SED properties correlate with colors. The color bins were chosen so that each subsample has roughly the same number of sources. This template library is ideal for applying to high 
redshift sources that only have IR photometry available. 
Although created from the same sources, this library differs from the MIR-based Library in part because $S_8/S_{3.6}$ is sensitive to dust obscuration which is an effect missed when separating by \fAGN alone. Furthermore, 
$S_{250}/S_{24}$ is sensitive to dust temperature, and when we sort sources by this color, we can test how strong the link is between the increase in warm dust and \fAGN. This is subtly different from linking $T_w$ and \fAGN in the MIR-based Library. 
In essence, with the MIR-based Library, we sorted sources by mid-IR AGN emission and looked for trends with the far-IR. Here, we begin by separating according to a far-/mid-IR color and test whether we recover the same trends with \fAGN.

In \citet{kirkpatrick2013}, we defined an empirical separation between AGN and SFGs
\begin{equation}
\log(S_8/S_{3.6})=0.74\times\log(S_{250}/S_{24})-0.78
\end{equation}
shown as the dark line in Figure \ref{fig:colors}. By dividing colorspace into eight quadrants, we can refine this AGN selection technique to include composites as well. Our color criteria can be used to estimate \fAGN of a source, and for this purpose we list the mean \fAGN in each color region in Table \ref{tbl:colors}. 
The upper three quadrants, COLOR6, COLOR7, and COLOR8, have the smallest spread of \fAGN, so these color criteria are excellent for selecting strong AGN sources. SFGs are confined to the lower three quadrants, COLOR1, COLOR2, and COLOR3. The middle regions, COLOR4 and COLOR5, have a large population of composite galaxies, which show strong star forming and AGN signatures.  

Figure \ref{fig:colors} and Table \ref{tbl:colors} demonstrate that there is a large spread in the observed colors of AGN due to differing levels of dust obscuration, varying amounts of dust heating by the AGN, or slight differences in the intrinsic SED of the host galaxy \citep{mullaney2011}. We have tested potential effects of heavy obscuration using the high $\tau$ AGN template from \citet{sajina2012} and find that obscuration can account for some of the scatter in the COLOR6, COLOR7, and COLOR8 quadrants, but it will not cause an AGN to mimic the colors of an SFG or composite. The spread in our AGN SEDs is consistent with what is observed in the local Universe, where local LIRGs with a significant mid-IR AGN contribution have a larger range of silicate absorption and PAH emission strengths compared with star forming LIRGs \citep{stierwalt2014}.

\begin{deluxetable}{l c c c c}
\tablecolumns{5}
\tablecaption{Mid-IR AGN Strength of IR Color Regions \label{tbl:colors}}
\tablehead{\colhead{Region\tablenotemark{a}} & \colhead{$S_8/S_{3.6}$\tablenotemark{b}} & \colhead{$S_{250}/S_{24}$\tablenotemark{b}} & \colhead{Mean \fAGN} & \colhead{$\sigma$\tablenotemark{c}}}
\startdata
COLOR1 & $<1.3$	& $\ge 35$	& 0.20	& 0.24 \\
COLOR2 & $<1.3$	& $<35$		& 0.26	& 0.27  \\
COLOR3 & 1.3-2.5	& $\ge 20$	& 0.41	& 0.34 \\
COLOR4 & 1.3-2.5	& $<20$		& 0.70	& 0.26  \\
COLOR5 & 2.5-5.0	& $\ge 13.5$	& 0.75	& 0.27 \\
COLOR6 & 2.5-5.0	& $<13.5$		& 0.94	& 0.06 \\
COLOR7 & $\ge 5.0$	& $\ge 14.5$	& 0.83	& 0.12 \ \\
COLOR8 & $\ge 5.0$	& $<14.5$		& 0.91	& 0.10
\enddata
\tablenotetext{a}{Each region is illustrated in the bottom panel of Figure \ref{fig:colors}.}
\tablenotetext{b}{The color limits corresponding to each region.}
\tablenotetext{c}{The standard deviation of \fAGN around the mean values so that the reader can understand the typical spread of \fAGN for each region.}
\end{deluxetable}

\begin{figure*}
\centering
\includegraphics[width=7in]{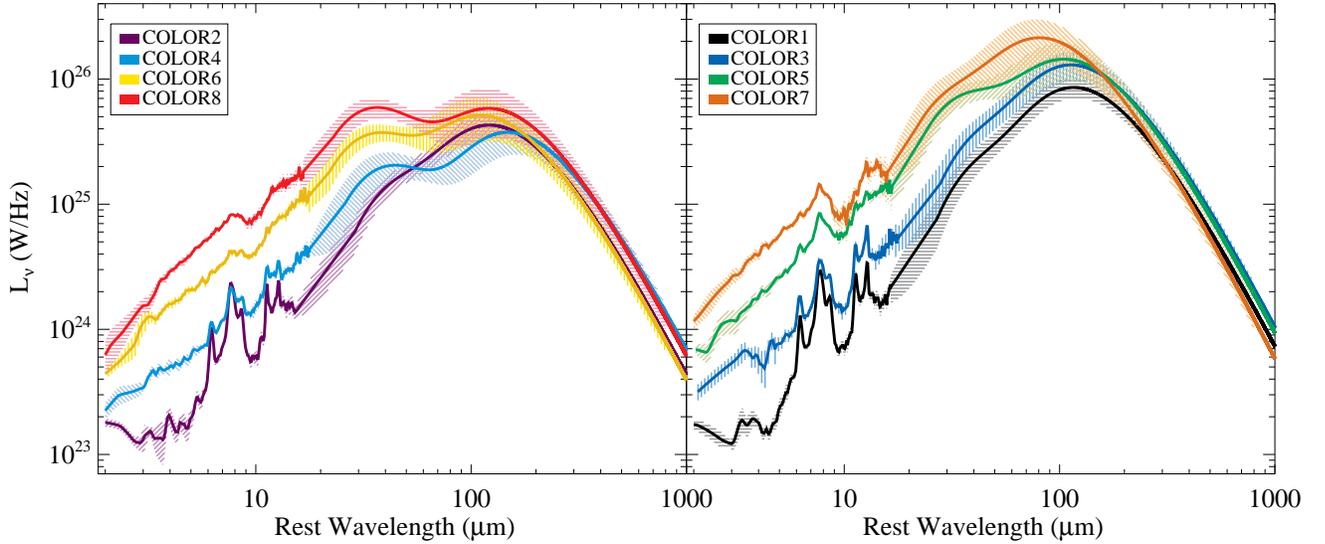}
\caption{Color-based Template Library. The sources have been selected by their location in IR colorspace. We have separated the eight templates into two panels to allow for easier comparison based on their far-IR colors. The templates in the left panel have lower $S_{250}/S_{24}$ ratios and a stronger warm dust contribution than the templates in the right panel.\label{fig:col_temp}}
\end{figure*}

We show all eight templates in Figure \ref{fig:col_temp}, where we have separated the templates into two panels for easier comparison based on $S_{250}/S_{24}$ color divisions. For consistency with the other libraries, we also truncate these templates below 2\,$\mu$m, although the observed 3.6\,$\mu$m photometry point falls below this threshold at $z>0.8$. In general, there is a lot of scatter below 2\,$\mu$m, and since we are not fitting this regime with any physical model, we truncate the templates to avoid over-interpretting the data.
The templates in the right panel, with higher $S_{250}/S_{24}$ ratios, all have clearly visible PAH features. 
In the left panel, the warm dust component is clearly prominent, and the near-IR slope grows steeper as the AGN becomes stronger. 
The COLOR8 template still has larger errors around the cold dust component than the other templates, which is primarily attributable to 
selection effects. This template is comprised of the strongest AGN sources, and these sources typically lie at higher redshift, producing less 
photometry in the Rayleigh-Jeans tail (at the median redshift, 500\,$\mu$m observed frame corresponds to $\sim170\,\mu$m rest frame). The cold dust emission of this template agrees with the available submillimeter observations (Appendix \ref{app:temp}), but since submillimeter observations are necessarily biased towards colder sources, we caution against using this template to extrapolate to submillimeter wavelengths.

\begin{figure}
\centering
\includegraphics[width=3.3in]{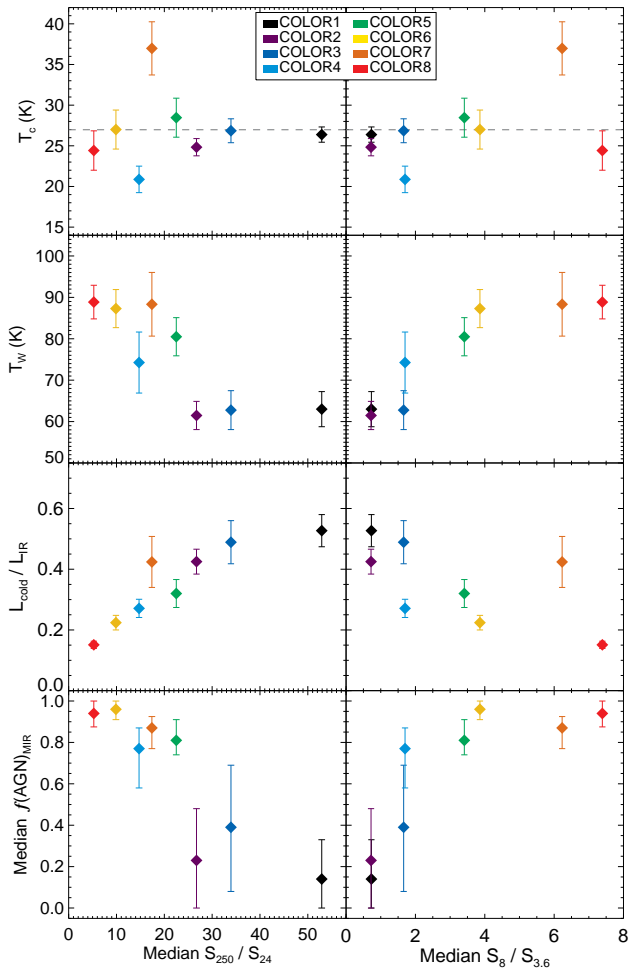}
\caption{Dust properties of the Color-based Templates. Warm dust temperature, $L_{\rm cold}/L_{\rm IR}$, and \fAGN are a strong function of the two IR colors, due to the fact that $T_w$ and $L_{\rm cold}/L_{\rm IR}$ are influenced by heating from the growing AGN. In contrast, the cold dust temperature comes from the host galaxy, and this property is insensitive to IR color. \label{fig:dust_color}}
\end{figure}

We explore the dust properties as a function of the IR colors in Figure \ref{fig:dust_color}. By grouping sources based on observed properties, we are able to look for correlations between observed properties and intrinsic properties such as dust temperature and \fAGN.
 The cold dust temperature shows no obvious correlation with $S_{250}/S_{24}$ and $S_{8}/S_{3.6}$. This is similar to what we observe for the MIR-based Library. 
$S_{250}/S_{24}$ is correlated with $T_w$ and $L_{\rm cold}/L_{\rm IR}$.
The trend with $S_{250}/S_{24}$ is expected since this ratio covers the wavelength range that we fit the warm MBB. The trend with $L_{\rm cold}/L_{\rm IR}$ clearly demonstrates that $S_{250}/S_{24}$ is a good proxy for the relative amount of cold dust emission by a galaxy.

We are also able to observe the effect of the AGN on multiple portions of the IR SED when we examine the trends between $L_{\rm cold}/L_{\rm IR}$, $T_w$ and the near-IR color $S_8/S_{3.6}$. The prominence and temperature of the warm dust component increases as this color increases. The increase of $S_8/S_{3.6}$ is due to dust heating from the torus outshining the stellar bump, producing a power-law whose slope depends on the amount of dust extinction. The far-IR emission is not necessarily occurring on the same spatial scales as the near-IR emission, since dust at different temperatures is required to produce emission in each wavelength range. The correlation between $L_{\rm cold}/L_{\rm IR}$, $T_w$ and $S_8/S_{3.6}$ could indicate that the same mechanism is responsible for both far-IR and near-IR sources of dust heating. We can test whether an AGN or star formation is the primary driver of the warm dust temperature by comparing the warm dust temperature with the amount of $L_{\rm IR}$ due to star formation or an AGN (calculated in Section \ref{sec:five}). We find no correlation between $T_w$ and $L_{\rm IR}^{\rm SF}$ (listed in Table \ref{tbl:temp_stats}), but we see a strong relationship between $T_w$ and $L_{\rm IR}^{\rm AGN}$, hinting that AGN luminosity is responsible for the increase in the warm dust temperature. This AGN-heated warm dust cannot be directly associated with the torus, which is on much smaller spatial scales and much hotter, but is most likely AGN-heated dust in the host galaxy.

The bottom row of Figure \ref{fig:dust_color} shows the trends between \fAGN and each color. We plot the median \fAGN and include the upper and lower quartiles to illustrate the spread in each subsample. \fAGN is strongly correlated with $S_{250}/S_{24}$, illustrating that the ratio of far- to mid-IR emission is an excellent indicator of mid-IR AGN strength. On the other hand, \fAGN shows a bimodality with  $S_8/S_{3.6}$ rather than a linear trend. Sources with \fAGN$>$0.8 have a range of $S_8/S_{3.6}$ values, partially explained by differing extinction levels, while sources with \fAGN$<0.2$ have $S_8/S_{3.6}\leq1$.

\subsection{Comprehensive Templates}
\label{sec:com_temp}


\begin{figure*}
\centering
\includegraphics[width=7in]{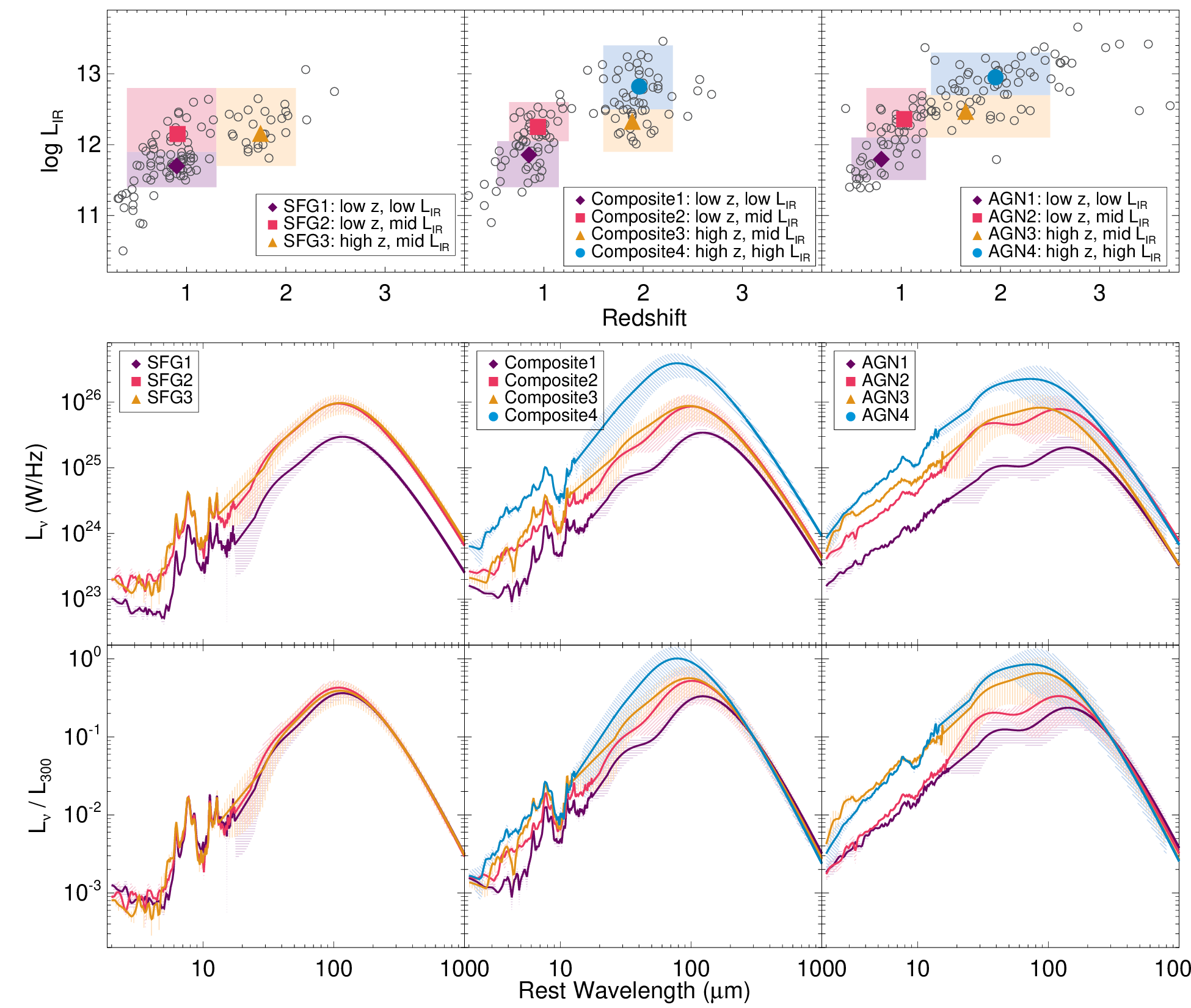}
\caption{We present our library of Comprehensive Templates where we have separated sources by \fAGN, redshift, and $L_{\rm IR}$. SFGs (\fAGN$<$0.2) are plotted in the left row, Composites (\fAGN=0.2-0.8) in the middle row, and AGN (\fAGN$>$0.8) in the right row. The top row shows our selection criteria as the shaded regions for the sources that comprise each template. We overplot the median $L_{\rm IR}$ and $z$ in each subsample (symbols here correspond to symbols used in Figure \ref{fig:lza_stats}). Shaded regions were selected to maximize completeness while optimizing the median redshift and $L_{\rm IR}$ so that we can compare templates at similar redshifts and $L_{\rm IR}$. The middle row shows the library of templates at the intrinsic luminosity density of each template, while in the bottom row, the templates have all been normalized at 300\,$\mu$m to allow easier comparison of the mid- and far-IR features. The large uncertainties on the far-IR emission of the AGN templates are due to the intrinsic scatter of SED shapes among these sources and a lack of data constraining the Rayleigh-Jeans tail due to the redshifts of the sources. The shape of the SFGs is remarkably consistent. The peak of the SED increases steadily for the Composites, as can clearly be seen in the bottom middle panel. The AGN1 and AGN2 templates, which are created from the lower redshift AGN, have a distinctly different far-IR shape than the AGN3 and AGN4 templates, indicating a possible temperature evolution with redshift. \label{fig:lza_temp}}
\end{figure*}

In this library, we look for evolution of the IR SED shape as a function of AGN strength, redshift, and $L_{\rm IR}$.
We initially separate our sources by \fAGN. We have three categories: (1) SFGs (\fAGN$\leq$0.2); (2) Composites (\fAGN=0.2-0.8); (3) AGN (\fAGN$\geq$0.8). 
These categories are motivated by the trends we seen in the MIR-based and Color-based Libraries. The bottom panels of Figure \ref{fig:dust_color} demonstrate that the near-IR and far-IR colors change significantly when \fAGN$>0.8$ and \fAGN$<0.2$, making these natural selection thresholds.

Within these \fAGN categories, we further separate by redshift and by $L_{\rm IR}$, selected to maximize completeness in each $L_{\rm IR}-z$ bin. We have optimized the redshift and $L_{\rm IR}$ selection criteria in order to 
have at least
two subsamples with similar median redshifts and two subsamples with similar $L_{\rm IR}$ values, so that we can examine the shape of the templates as a function of both redshift and $L_{\rm IR}$. The division of the sources is illustrated in the top rows of Figure \ref{fig:lza_temp}. 
Our Comprehensive Library is shown in the middle rows of Figure \ref{fig:lza_temp}, and in the bottom rows we have normalized the templates at 300$\,\mu$m to allow for easier comparison. The large uncertainties in the range 15-30\,$\mu$m for the higher redshift templates is due to a scarcity of photometric data, particularly for the SFGs due to these sources being intrinsically fainter in this regime. The high redshift Composites (blue  and gold templates) have large far-IR errors as a result of few sources in this subsample. In contrast, the large uncertainties on the far-IR for the AGN templates are caused by the intrinsic scatter of SED shapes among these subsamples and a lack of data constraining the Rayleigh-Jeans tail for the high redshift templates. 

The bottom panels of Figure \ref{fig:lza_temp} qualitatively illustrate a fundamental difference between our SFGs and our Composites and AGN. The SFGs have a high degree of similarity between all
$L_{\rm IR}$s and redshifts. Our analysis suggests that on average, the SFG SED does not evolve with redshift or luminosity for these types of massive dusty galaxies. Any evolution between $L_{\rm IR}$ and dust temperature is driven not by an intrinsic change in the ISM of high redshift (U)LIRGs, but by a different process, such as a growing AGN. We should note that our selection criteria is biased towards sources with strong PAH emission (which we will comment on further in Section \ref{sec:six}), and we are examining only one order of magnitude in $L_{\rm IR}$ which is possibly too narrow of a range to expect to see any strong trend between $T_c$ and $L_{\rm IR}$.

\begin{figure}
\centering
\includegraphics[width=3.4in]{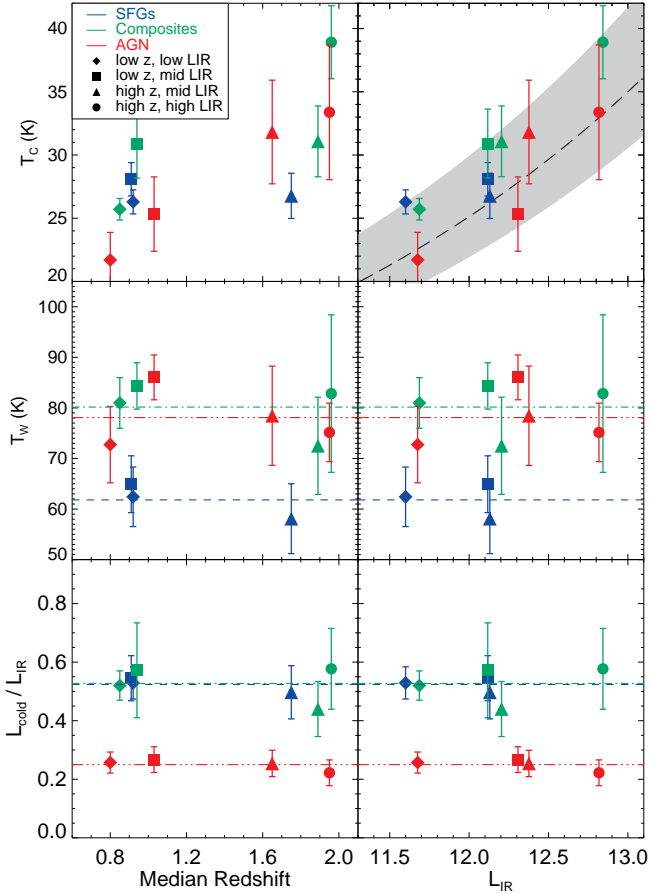}
\caption{We plot the derived cold and warm dust temperatures ($T_c, T_w$) for each of our 11 Comprehensive Templates, as well as the fraction of $L_{\rm IR}$ due to the cold dust component. We shade the points according to \fAGN, so that SFGs are blue, Composites are green, and AGN are red; in addition, symbols correspond to redshift and $L_{\rm IR}$ of the templates (illustrated in the top row of Figure \ref{fig:lza_temp}). Here, we plot the far-IR dust parameters as a function of the median redshift of the sources used to create each template and as a function of the template $L_{\rm IR}$.
In the lower panels, we plot the median $T_w$ and $L_{\rm cold}/L_{\rm IR}$ for the SFGs, Composites, and AGN templates as the dashed lines. Cold dust temperature increases with $L_{\rm IR}$ and redshift for the Composites and AGN, while the SFGs show no evolution. $L_{\rm cold}/L_{\rm IR}$ is significantly lower in the AGN templates because dust heated by the AGN is now outshining much of the dust heated by starlight alone. In the upper right panel, we have plotted the $L_{\rm IR}$-T relation from \citet{casey2012a}. \label{fig:lza_stats}}
\end{figure}

We quantify the far-IR dust properties of the Comprehensive Library in Figure \ref{fig:lza_stats}. 
In the left column, we plot $T_c$, $T_w$, and $L_{\rm cold}/L_{\rm IR}$ as a function of the median redshift of the sources that were used to create each template. In the right column, we plot these properties as a function of template $L_{\rm IR}$. 
The three SFG templates all have the same $T_c$, $T_w$, and $L_{\rm cold}/L_{\rm IR}$ regardless of $L_{\rm IR}$ or redshift, effectively demonstrating the lack of evolution in these sources, on average. 

In contrast, the Composites and AGN show a clear increase in $T_c$, also evident in Figure \ref{fig:lza_temp} where the peak of the SED shifts with increasing $L_{\rm IR}$ and $z$. For the Composites, the increase in $T_c$ is correlated with $L_{\rm IR}$, as can be seen clearly by examining the green points in the top right panel of Figure \ref{fig:lza_stats}. The Composite2 and Composite4 templates (square and circle) each have $T_c$ at least 5\,K higher than the Composite1 and Composite3 templates (diamond and triangle), despite lying at similar redshifts, respectively. In contrast, for the AGN, the increase of $T_c$ is more strongly correlated with redshift, although $T_c$ has large uncertainties. It is important
not to overstate this distinction, since our high redshift subsamples are biased towards sources with high $L_{\rm IR}$. Nevertheless, the clear difference between the AGN and Composites hint that different mechanisms could be driving
the evolution of dust temperature. For the Composites, this change could be driven by an increase in the importance of mergers which produce compact starbursts and warmer dust \citep{armus2007}, while for the AGN, there might be an intrinsic evolution in the effect of an AGN on the IR SED with redshift, as galaxies have higher gas fractions and clumpier ISMs. We have morphological classifications for our xFLS sample \citep{zam2011}, and these data hint at an increase in the number of interacting galaxies for the higher luminosity Composite templates, but not for the AGN. We will examine the individual morphologies of our galaxies in a future work.

The relationship between dust temperature and $L_{\rm IR}$ that we see for the Composites and AGN is consistent with what is derived for a large sample of high redshift SPIRE 250\,$\mu$m selected galaxies \citep{casey2012a}. We have over plotted the relation from \citet{casey2012a} as the black line and grey shaded region in the top right panel. \citet{casey2012a} make no attempt to separate sources into SFGs or AGN. The increase in $T_c$ in our Composite and AGN templates is noticeably absent in our SFG templates, implying that the $L_{\rm IR}-T$ relation is not driven by a simple change in the ISM with redshift. 
However, we lack a higher luminosity SFG template, so it is impossible to say conclusively that we would not observe a trend among the SFGs if we had full coverage at high $L_{\rm IR}$.

The middle panels of Figure \ref{fig:lza_stats} show the warm dust temperatures. We plot the average warm dust temperatures for the SFG, Composite, and AGN templates as the dashed lines to allow for easier comparison.
$T_w$ does not evolve strongly with either $L_{\rm IR}$ or redshift. Furthermore, $T_w$ is similar for the Composite templates (the average $T_w$ for all four templates is 80\,K) and AGN templates (average $T_w=78\,$K), while SFGs have a lower average $T_w$ of 62\,K. This suggests that the mechanism responsible for heating the warm dust component is linked with \fAGN, and it also confirms that our threshold of \fAGN$<$0.20 for selecting SFGs is well-founded. The middle panels are best interpreted in comparison with the bottom panels. When we examine the fraction of $L_{\rm IR}$ due to the cold dust component, we find that the Composites and SFGs have similar fractions ($L_{\rm cold}/L_{\rm IR}\sim0.52$), while the AGN only have $L_{\rm cold}/L_{\rm IR}\sim0.25$, despite having similar warm dust temperatures as the Composites. 

Taken together, these data present a picture whereby the Composites are a true mix of the SFGs and AGN. 
In SFGs, the warm dust arises from star-forming regions. Once an AGN significantly contributes to the mid-IR emission, warm dust heated by the AGN becomes more luminous than the warm dust located in star forming regions, producing an increase in $T_w$. However, although a significant amount of the dust is now heated by a central AGN in the Composite galaxies, the cold dust emission still dominates $L_{\rm IR}$. For the AGN, a larger fraction of the dust mass is heated by the central AGN as indicated by the lower $L_{\rm cold}/L_{\rm IR}$ ratios. 

\section{Relation between Mid-IR and Far-IR}
\label{sec:five}
\begin{figure*}
\centering
\includegraphics[width=6.9in]{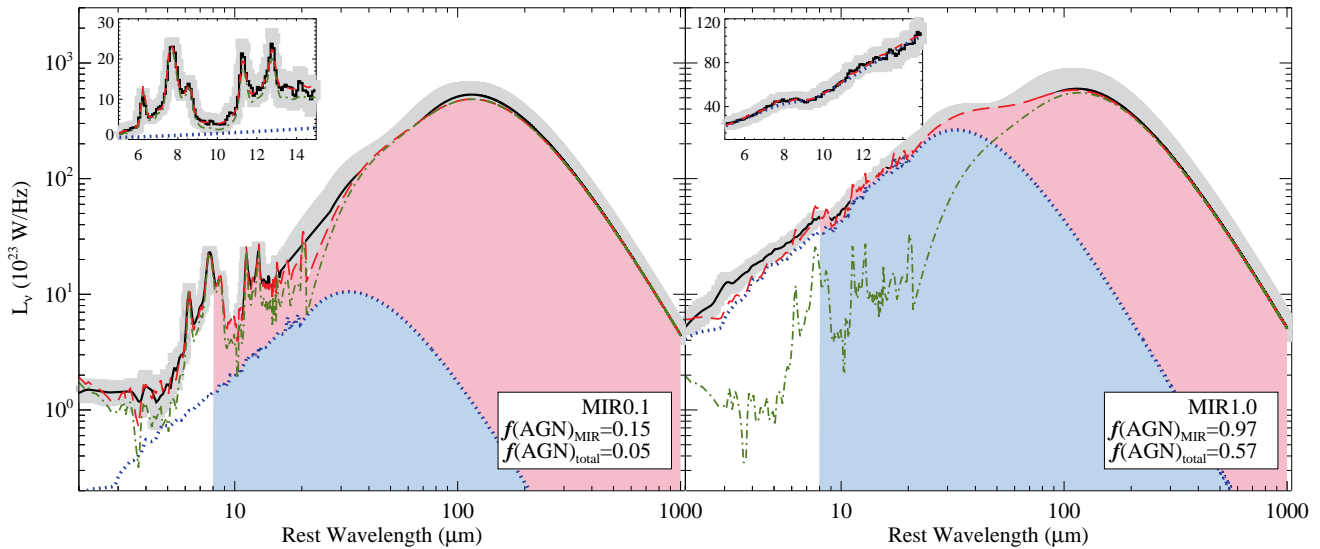}
\caption{We illustrate our full IR decomposition technique for a star forming template (MIR0.1, left) and an AGN template (MIR1.0, right). We find a best fit model (red dashed line) by simultaneously fitting the $z\sim1$ Star Forming SED (green dot-dashed line) and Featureless AGN SED (blue dotted line), with extinction if required, from \citet{kirkpatrick2012}. We then integrate under the AGN component (blue dotted line) to calculate \fTOT which is the fraction of $L_{\rm IR}$(8-1000\,$\mu$m) due to AGN heating. We have illustrated the integrated portion of the model and AGN component with the shaded regions. In the insets, we show the mid-IR decomposition (Equation \ref{eq:decomp}), used to calculate \fAGN from 5-15\,$\mu$m. \label{fig:decomp}}
\end{figure*}

Throughout this paper, we have been discussing the AGN strength in the mid-IR. We now wish to explore how \fAGN correlates with the total contribution of the AGN to $L_{\rm IR}$. We use the three template libraries presented in Section 4 to calculate a conversion between \fAGN and \fTOT.
First, we perform mid-IR spectral decomposition, described in Section \ref{sec:decomp}, on each template from $\lambda=5-15\,\mu$m. Next, we use a similar decomposition technique for the entire template. We fit simultaneously the $z\sim1$ Star Forming SED and the Featureless AGN SED from \citet{kirkpatrick2012}. We  removed the cold dust component from the Featureless AGN SED, as this component arises from the host galaxy. We have verified that the remaining SED does not contain any host emission using the decomposition package {\sc decompir} presented in \citet{mullaney2011}. We modify the Featureless AGN SED with the extinction curve from \citet{draine2003}, where we hold $\tau$ fixed to the values derived from the mid-IR spectral decomposition. The normalizations of the AGN and Star Forming SEDs are the only free parameters, and we allow them to vary simultaneously. Figure \ref{fig:decomp} illustrates two decomposition examples. We calculate \fTOT by integrating under the Featureless AGN SED to obtain the $L_{\rm IR}$ of the AGN component, and we express this as a fraction of the total $L_{\rm IR} (8-1000\,\mu$m). This simple decomposition technique works well for 80\% of our templates. However, the Composite4, AGN1, AGN4, MIR6, COLOR4, and COLOR7 templates are not well fit, resulting in poor $\chi^2$, due to the cold dust peaking at significantly warmer temperatures than the $z\sim1$ Star Forming SED, and we exclude them from the analysis below.

\begin{figure}
\centering
\includegraphics[width=3.4in]{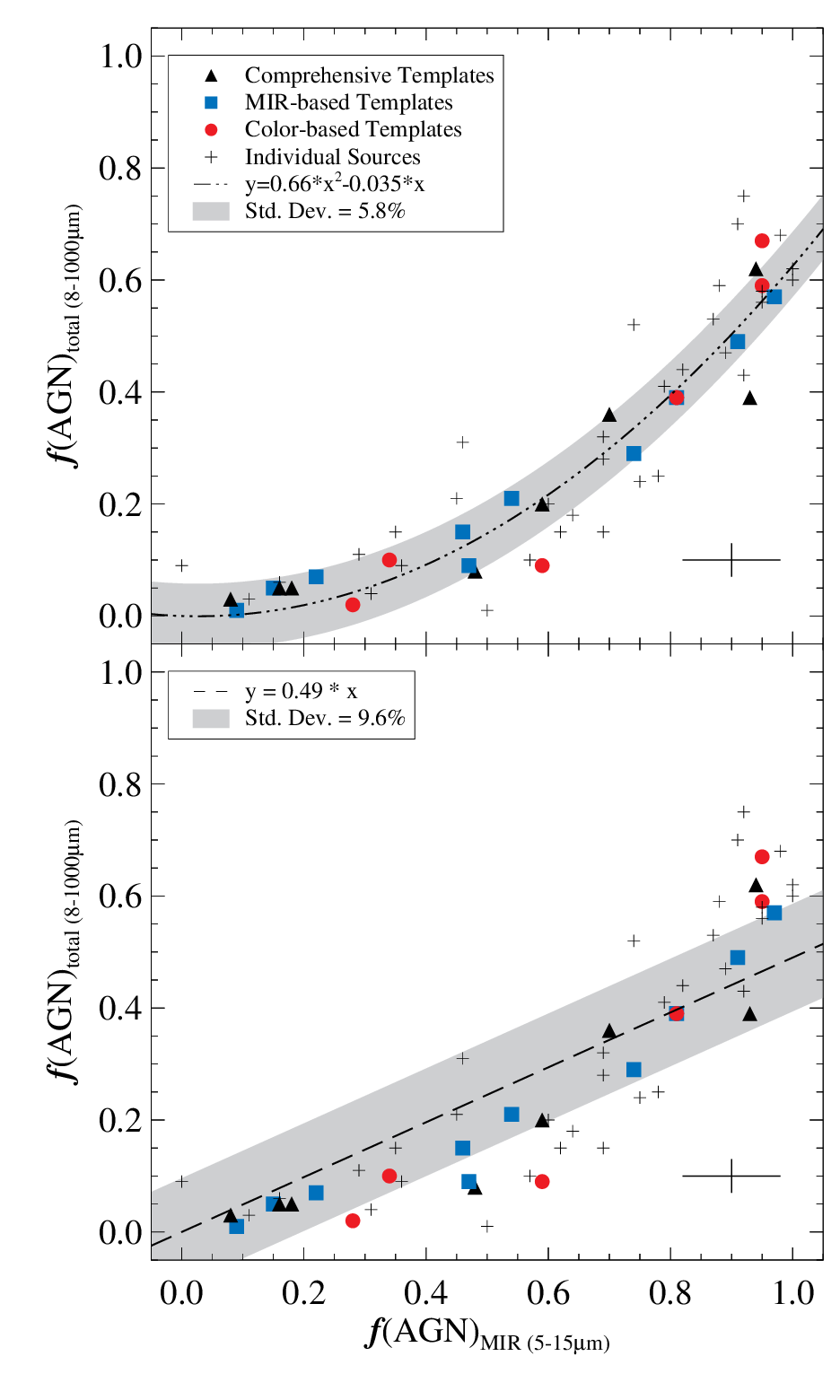}
\caption{\fAGN v. \fTOT. Typical uncertainties on each parameter are shown in the lower right corner. \fAGN has a higher error because we have included an extinction component in this fit, while when fitting \fTOT, we only allow the relative normalizations to vary. We fit a quadratic equation ({\it top panel}) and a linear relation ({\it bottom panel}) to our templates (filled symbols). The standard deviation of the templates around each relation is shown as the grey shaded region. We have a handful of sources ($\sim30$) with exceptionally well-sampled SEDs, allowing us to decompose the entire SED into an AGN and star forming component. We plot these sources as the crosses. They are not included in the linear or quadratic fits or the standard deviation calculations, but they agree remarkably well with these relations, ensuring the reliability of our results. The quadratic relation provides a better fit to the data. \label{fig:mir_tot}}
\end{figure}

We plot \fAGN versus \fTOT in Figure \ref{fig:mir_tot}, where the filled symbols correspond to the template libraries. In the bottom panel, we quantify the relationship between \fAGN and \fTOT with a simple linear scaling (plotted as the dashed line)
\begin{equation}
\fTOT=(0.49\pm0.02)\times\fAGN
\end{equation}
where we have weighted each template by the number of sources comprising it. We require the linear fit to have a $y$-intercept of 0, so that \fTOT=0 when \fAGN=0. The standard deviation around this relation is 9.6\% (grey shaded region). 
Our composite templates lie below the dashed line while the AGN lie above it, indicating that a simple linear scaling may not be the best choice, especially if \fAGN is well known.

In the top panel, we use a quadratic equation to quantify the relationship between \fAGN and \fTOT:
\begin{multline}
\label{eq:mir}
\fTOT = (0.66\pm0.09)\times \fAGN^2\\
-(0.035\pm0.07) \times\fAGN
\end{multline}
Again, we have weighted each template by the number of sources that comprise it, and we have forced the $y$-intercept to be 0.
We plot this relationship, and the corresponding standard deviation of 5.8\%, as the dark dashed line and grey shaded region, respectively. The smaller standard deviation indicates that this is a better fit for our templates. 

We independently test the validity of our relation using the SEDs of individual sources with exceptional photometric coverage of the far-IR. We have already decomposed the mid-IR spectra of these sources, and we decompose the full SEDs using the same procedure as for our templates.
We plot these sources as the black crosses. The scatter among the individual sources better illustrates the uncertainty associated with our decomposition methods, but in general, the trend between \fAGN and \fTOT for the sources agrees remarkably well with the templates and reinforces that the quadratic relation is a better fit to the data than the linear scaling.

The quadratic relation arises 
because AGN-heated dust emission falls off sharply after 40\,$\mu$m, so $L_{\rm IR}$ is dominated by stellar heating after this wavelength. That is, until the AGN boosts the warm and hot dust emission enough to outshine the cold diffuse dust which Figure \ref{fig:mir_stats} indicates happens when \fAGN$>$0.70. This non-linear relationship supports what we found earlier, that composites are an intermediate class between SFGs and AGN where most of the far-IR emission can be attributed to star formation, although the AGN is dominating at shorter wavelengths.

Since we have determined \fTOT, we can scale $L_{\rm IR}$ by (1-\fTOT) to obtain $L_{\rm IR}^{\rm SF}$, the portion of $L_{\rm IR}$ due only to heating by stellar radiation (listed in Table \ref{tbl:temp_stats}). 
We could not decompose the full IR SED of the Composite4, AGN1, AGN4, MIR6, COLOR4, and COLOR7 templates, so we calculate \fTOT using Equation \ref{eq:mir}. 
$L_{\rm IR}^{\rm SF}$ is a crucial quantity for obtaining accurate star formation rates. In future surveys, particularly with JWST, \fAGN can be determined using mid-IR spectroscopy; this can then be converted into \fTOT using one of our relations, and $L_{\rm IR}$ can be scaled accordingly so as not to overestimate star formation rates. Carefully removing AGN contribution to $L_{\rm IR}$ will provide a more accurate understanding of the buildup of stellar mass in the early Universe.

Finally, we wish to comment on how our far-IR properties relate to another commonly used measure, $L({\rm FIR})$ which is the integrated luminosity from $50-300\,\mu$m. $L_{\rm cold}$ accounts
for most of the emission in this wavelength regime, and we find a nearly linear relationship between $L_{\rm cold}$ and $L({\rm FIR})$:
\begin{equation}
L({\rm FIR})=6.92\, L_{\rm cold}^{0.94}
\end{equation}
$L_{\rm cold}$ also accounts for the bulk of $L_{\rm IR}$ attributed to star formation by our decomposition technique, and $L_{\rm cold}$ and $L_{\rm IR}^{\rm SF}$ have a nearly linear relationship as well
\begin{equation}
L_{\rm IR}^{\rm SF}=9.31\, L_{\rm cold}^{0.94}
\end{equation}
The strong orrelation between $L({\rm FIR})$ and $L_{\rm IR}^{\rm SF}$ strengthens our conclusion that heating by star formation accounts for the bulk of the cold, far-IR emission.

\section{Discussion}
\label{sec:six}
\subsection{Consistency in Star Forming Galaxies over Cosmic Time}
We have carefully decomposed the mid-IR spectra of our sources, allowing us to classify galaxies harboring a buried AGN that may not be visible at other wavelengths. 
An additional benefit of this classification scheme is that it enables us to isolate the mid- and far-IR properties of purely star forming galaxies over a large range in redshift. 
Within the Comprehensive Library, we have determined the average SEDs of pure SFGs with median redshifts of $z\sim0.8$ and $z\sim1.7$, and these templates are indistinguishable (left column of Figure \ref{fig:lza_temp}). In particular, we find that mid-IR classification is an excellent predictor of far-IR emission, and this is not a trivial result since mid-IR and far-IR emission are tracing different dust populations at different spatial locations. 

Before we discuss the full IR SED, we want to comment on our mid-IR identification technique. Our technique for selecting SFGs hinges on the PAH emission, and it is possible that we could misidentify SFGs that have mid-IR emission dominated by star formation, but also have weak PAH emission. However, based on previous results in the literature, we do not think these types of galaxies are common in samples of (U)LIRGs; locally, such galaxies are low-metallicity dwarfs \citep[e.g.,][]{wu2006}. The similarity of PAH emission among star forming galaxies is also seen in \citet{battisti2015}, which compares PAH features of local star forming galaxies ($L_{\rm IR}\sim10^{10}\,L_\odot$) and finds remarkable consistency. These galaxies have been classified as SFGs according to optical emission line ratios. \citet{battisti2015} also compares the average PAH emission of these local SFGs with the star forming templates from \citet{kirkpatrick2012} and find they are consistent, showing no evolution of PAH emission features with redshift or luminosity. \citet{petric2011} classifies local (U)LIRGs from the Great Observatories All Sky Survey \citep[GOALS;][]{armus2009}. as SFGs based on mid-IR flux ratios and finds that PAH emission in all SFGs is qualitatively similar. On the other hand, Polletta et al. (2008) and Bauer et al. (2010) conclude that the mid-IR continuum in high redshift ULIRGs with weak PAH features is dominated by quasar emission, although these sources can still have a significant amount of $L_{\rm IR}$ due to star formation. Our results are consistent with these previous studies.

Our SFG templates have no significant change in $T_w$, $T_c$ or $L_{\rm cold}/L_{\rm IR}$ with redshift or $L_{\rm IR}$, effectively demonstrating that the average dust heating in SFGs remains constant over a broad epoch. 
This result does not contradict observations from \citet{bethermin2015}. For a sample of main sequence galaxies spanning a redshift range of $z=0.5-4$, the authors conclude that the average interstellar radiation field, measured by the parameter $\langle U\rangle$, increases as $\langle U\rangle \propto (1+z)^{1.15}$. $\langle U \rangle$ is proportional to dust temperature, indicating that the dust temperature should be increasing and the peak of the SED should be shifting to shorter wavelengths. 
However, when we plot the stacked detections in \citet{bethermin2015} in the same redshift range as our templates ($0.25<z<1.25$ and $1.25<z<2.00$), we find that the stacked fluxes are consistent with our SEDs within the uncertainties. The evolution in SED peak observed in \citet{bethermin2015} is not strong enough to be evident over the redshift range we are probing, which is more limited than that study.

Our observed lack of SED evolution is
consistent with observations of the larger sample of GOODS-{\it Herschel} SFGs;
the ratio of  PAH to far-IR emission, as traced by $L_{\rm 8\,\mu m}/L_{\rm IR}$, is constant from $z=0-2.5$ providing evidence that the IR SEDs of normal, non-interacting, dusty SFGs do not evolve strongly \citep{elbaz2011}. 
In contrast, local ULIRGs have a deficit of PAH emission compared with less luminous, normal SFGs \citep[e.g.,][]{veilleux2009}.
However, at higher redshift, this deficit is seen to shift to higher $L_{\rm IR}$, so that high redshift ULIRGs have $L_{\rm PAH}/L_{\rm IR}$ ratios that mimic local LIRGs, indicating that local LIRGs might be an ideal comparison sample for our high redshift LIRGs and ULIRGs \citep{sajina2012,pope2013,stierwalt2013}.

\begin{figure}
\centering
\includegraphics[width=3.3in]{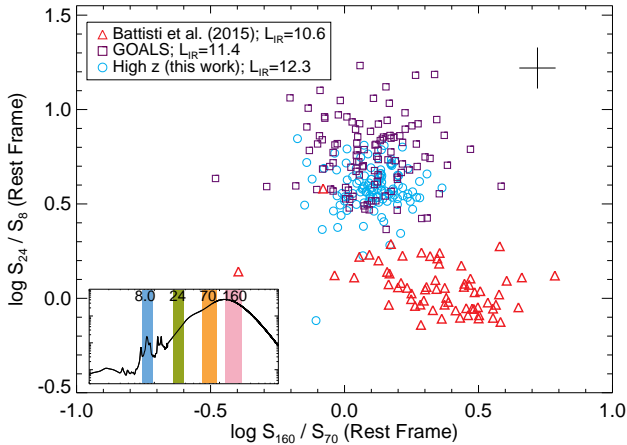}
\caption{We plot the colors $S_{160}/S_{70}$ (rest frame) and $S_{24}/S_8$ (rest frame) of three samples: local star forming LIRGs from GOALS (purple squares), local normal SFGs (red triangles), and our SFGs (blue circles). The median $L_{\rm IR}$ of each sample is listed in the legend. We have estimated the rest frame colors for our SFGs by sampling the MIR0.0-MIR0.2 templates within the template uncertainties. Average uncertainties produced by this method are illustrated by the cross in the upper right corner. Our high $z$ SFGs show a strong overlap with the GOALS LIRGs, but the less luminous local SFGs lie in a different region of colorspace. This indicates that the SEDs of local LIRGs are similar to the SEDs of high $z$ LIRGs and ULIRGs. \label{fig:local_colors}}
\end{figure}

We compare the IR colors of our high $z$ SFG sources with the observed frame IR colors of local LIRGs in Figure \ref{fig:local_colors}. We plot $S_{160}/S_{70}$ and $S_{24}/S_8$ for LIRGs from GOALS. The individual galaxies in GOALS all have mid-IR spectroscopy available allowing us to classify their mid-IR AGN emission using the same technique as for our high redshift galaxies. In Figure \ref{fig:local_colors}, we are only comparing mid-IR identified SFGs (\fAGN$<$0.2) in the GOALS and high $z$ samples. For the high $z$ sources, we have estimated rest frame $S_{160}/S_{70}$ and $S_{24}/S_8$ colors using a Monte Carlo technique to sample the MIR0.0, MIR0.1, and MIR0.2 templates within the template uncertainties at 8, 24, 70, and 160\,$\mu$m. We show the typical uncertainty on this synthetic photometry in the upper right corner. We also demonstrate the portion of the SED traced by these colors in the lower left corner. 

There is a strong overlap between our high $z$ SFGs and the GOALS SFGs. For comparison, we plot local less luminous ($L_{\rm IR}\sim10^{9}-10^{10}$) SFGs identified through optical emission line ratios \citep{odowd2011,battisti2015}. Although a few of these sources lie in the same region as GOALS and our SFGs, in general, these sources lie below and to the right of the more luminous galaxies. $S_{160}/S_{70}$ traces the peak of the SED, while $S_{24}/S_8$ traces the amount of warm dust relative to the PAH emission. That we see little difference in either of these colors between the GOALS and high $z$ SFGs indicates that the average SED of high redshift LIRGs and ULIRGs is remarkably similar to local LIRGs. In other words, the SEDs of luminous dusty galaxies may not evolve strongly with redshift, if we do not consider the extreme cases of compact mergers. This result agrees with \citet{stierwalt2013}, where the authors compared the average mid-IR spectra of the GOALS LIRGs with the average mid-IR spectra of submillimeter galaxies (SMGs) from \citet{menendez2009}. \citet{stierwalt2013} find that when all mid-IR AGN contribution is removed, the remaining spectra of local star-forming LIRGs is identical to high $z$ star-forming SMGs. We will explore more comparisons between the GOALS survey and our high redshift sources in a future work (Kirkpatrick et al.\,(2016), in prep).

\subsection{Demographics in Color Space}
\label{sec:colors}
\begin{figure}
\centering
\includegraphics[width=3.6in]{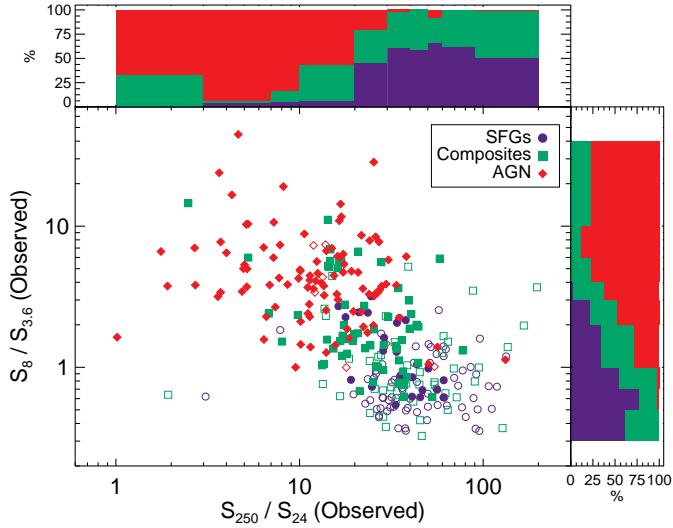}
\caption{Our sources in colorspace. We shade the sources according to their mid-IR power source. SFGs (blue circles) have \fAGN$<$0.2, composites (green squares) have \fAGN=0.2-0.8, and
AGN (red diamonds) have \fAGN$>$0.9. The full sample is plotted with the open symbols while sources with $S_{24}>0.9\,$mJy are plotted with the filled symbols. To the top and right, we plot the color demographics of our sources, where we have corrected for completeness effects by weighting the distributions by the 24\,$\mu$m number counts presented in \citet{bethermin2010}. These distributions are useful for estimating relative numbers of SFGs and AGN when only a couple of photometric data points exist for each source. \label{fig:col_dist}}
\end{figure}

Our templates represent the average SEDs of our sample, but the variation among sources in the sample can be seen in Figure \ref{fig:col_dist}, where we again plot the  colors $S_{250}/S_{24}$ and $S_8/S_{3.6}$ for our sample. We shade the sources according to \fAGN, 
so that SFGs (blue circles) have \fAGN$<$0.2, composites (green squares) have \fAGN=0.2-0.8, and AGN (red diamonds) have \fAGN$>$0.8.
Our sample is comprised of 24\,$\mu$m faint galaxies from GOODS and 24\,$\mu$m bright galaxies, primarily from xFLS. We illustrate the difference in these samples using filled and unfilled symbols. The filled symbols all have $S_{24}>0.9\,$mJy, and primarily lie to the upper left. AGN and SFGs lie in distinct regions in this colorspace, with AGN occupying primarily the upper left quadrant. 

Because we are combining sources selected at different $S_{24}$ thresholds, we do not have a complete sample. We account for completeness using the 24\,$\mu$m number counts from \citet{bethermin2010}. The authors list the number counts in 24\,$\mu$m flux bins from $S_{24}=0.035-100$\,mJy. We divide our sources into the same flux bins, and assign a weight to each source, so that our weighted number counts match what is presented in \citet{bethermin2010}. We then divide the colors $S_8/S_{3.6}$ and $S_{250}/S_{24}$ into refined bins and count the weighted number of sources in each bin to produce the color demographic histograms on the top and right of Figure \ref{fig:col_dist}. We list the color bins and weighted percentages of SFGs, composites, and AGN in each bin in Table \ref{tbl:dist}.
The composites are roughly equally distributed. This is because of variable levels of AGN within the composites but may also be linked to different triggering mechanisms for an AGN growth. Major mergers are known to produce warmer SEDs, but an AGN growing in a clumpy, extended disk likely has more cold dust \citep{elbaz2011}. AGN and SFGs separate cleanly in both $S_8/S_{3.6}$ and $S_{250}/S_{24}$, making each of these colors advantageous for selecting AGN and SFGs in a sample lacking spectroscopy or broad photometric coverage of the SED. 

We illustrate how our color demographics can be applied to large samples, to estimate the number of pure SFGs for example, using a catalog of 10,300 B$z$Ks with a detection in $S_8$ and $S_{3.6}$ \citep{lin2012}. We first determine the $S_8/S_{3.6}$ distribution of B$z$Ks using the bins listed in Table \ref{tbl:dist}. Then, we multiply the number of B$z$Ks in each bin by the respective percentages of SFGs in Table \ref{tbl:dist} to estimate the number of B$z$Ks that are SFGs. We calculate that from the full B$z$K catalog, only 23\% are pure SFGs. 
This could have implications for studies that see a redshift evolution in the shape of the SED. For example, using the same B$z$K catalog, \citet{magdis2012} measure 
$\langle U\rangle\propto(1+z)^{1.15}$ from $z=0-2$, where $U\propto T_c$. The authors have removed X-ray luminous AGN, but according to our $S_8/S_{3.6}$ color diagnostic, it is possible that many composites hosting an obscured AGN are included in their sample. We find a similar evolution of dust temperature with redshift for our Composites templates, but this is noticeably absent for our SFG templates (Figure \ref{fig:lza_stats}). Our color demographics can help estimate the level of contamination in a large sample from  galaxies that possess a mix of star formation and AGN activity. Our composites may be missed at X-ray wavelengths due to either high column densities or lower AGN X-ray luminosites. Moreover, the optical line ratios expected in composites is currently unconstrained at high redshift \citep{kartaltepe2015}. Our IR color technique then provides a unique opportunity to identify the AGN lurking in dusty, IR luminous galaxies.

\begin{figure}
\centering
\includegraphics[width=3.4in]{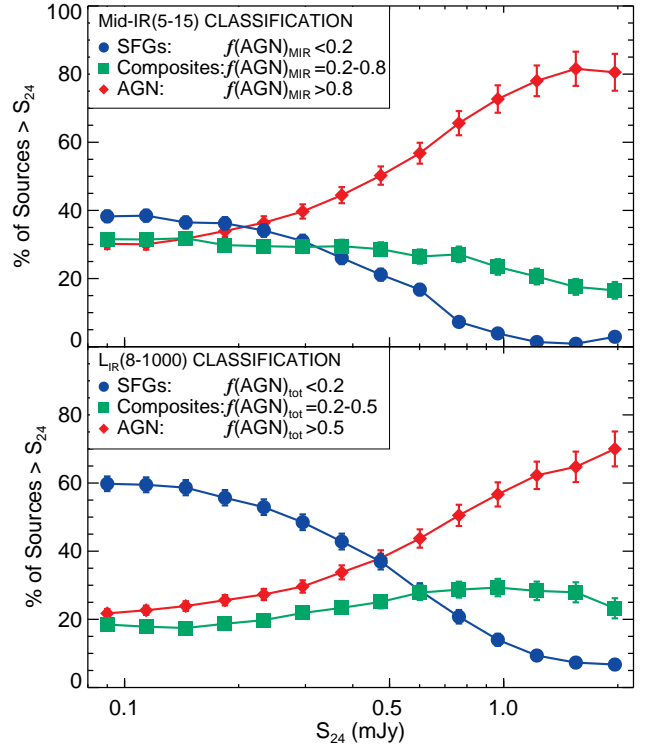}
\caption{We present the cumulative $S_{24}$ distribution of our sample. In the top panel, we classify sources as SFG, composite, or AGN based on \fAGN. In the bottom panel, we calculate \fTOT using Equation \ref{eq:mir} and then sort sources accordingly. Each point represents the percentage of SFGs, composites, or AGN brighter than a given 24\,$\mu$m flux. We have assigned weights to each galaxy based on it's 24\,$\mu$m flux density in order to reproduce the number counts presented in \citet{bethermin2010}. The cumulative distribution presented here is calculated using our sources' weights. We have attached Poisson errors to each point. AGN dominate at brighter fluxes. Composites contribute 20-30\% at all flux thresholds in both panels illustrating the necessity of properly estimating or removing composites from IR samples of SFGs. \label{fig:s24}}
\end{figure}

In Figure \ref{fig:s24}, we examine the effect that 24\,$\mu$m flux thresholds can have on the number of composites, SFGs, and AGN in a given sample. Again, we use the number count weights assigned to our sources, and plot the percentage of SFGs, composites, and AGN brighter than a given 24\,$\mu$m flux threshold (top panel). Throughout this paper, we have discussed SFGs, composites, and AGN using a mid-IR classification scheme, but in the bottom panel Figure \ref{fig:s24}, we classify sources according to \fTOT. Here, we have calculated \fTOT for all sources using Equation \ref{eq:mir}. SFGs have \fTOT$<$0.2, composites have \fTOT=0.2-0.5, and AGN have \fTOT$>$0.5, a threshold chosen because $L_{\rm IR}$ is now dominated by AGN emission. Although \fTOT is derived from \fAGN, we stress that the IR classifications are not the same as the mid-IR classifications.
Both panels of Figure \ref{fig:s24} show that 
AGN dominate at brighter fluxes ($S_{24}>0.5\,$mJy), so imposing a simple flux cut on a sample can easily remove large numbers of AGN. IR SFGs dominate the population when $S_{24}<0.4\,$mJy (bottom panel), but mid-IR SFGs never do (top panel). However, IR and mid-IR composite sources contribute about 20-30\% of a sample at all $S_{24}$ thresholds, so simply removing IR bright AGN or X-ray AGN does not account for all IR AGN emission. 

These demographics are useful for current and future high redshift studies, particularly with JWST. The MIRI instrument on JWST will have a broadband 25.5\,$\mu$m filter, so the $S_{24}$ distributions in Figure \ref{fig:s24} can inform desired sensitivities of a particular project.
Contamination by obscured AGN emission needs to be accounted for since Figure \ref{fig:s24} demonstrates that AGN and composites are non-negligible at all $S_{24}$ thresholds. Through MIRI and NIRcam, astronomers will also be able to obtain a color very similar to $S_8/S_{3.6}$, and so our color demographic in Figure \ref{fig:col_dist} and Table \ref{tbl:dist} can be used to estimate the number of SFGs and AGN in a given sample or select galaxies for further study.

\begin{deluxetable}{ccccc}
\tablecolumns{5}
\tablecaption{Color Distributions \label{tbl:dist}}
\tablehead{\colhead{Min Color} & \colhead{Max Color} & \colhead{SFG} & \colhead{Composite} & \colhead{AGN} }
\startdata
\cutinhead{$S_8/S_{3.6}$}
0.3	&	0.5	&	62.1	&	37.9	&	\, 0.0	\\
0.5	&	0.7	&	76.8	&	21.3	&	\, 1.8	\\
0.7	&	1.0	&	54.5	&	42.9	&	\, 2.6	\\
1.0	&	2.0	&	34.7	&	37.4	&	27.9	\\
2.0	&	3.0	&	21.9	&	30.2	&	47.8	\\
3.0	&	4.0	&	\, 2.2	&	36.0	&	61.8	\\
4.0	&	6.0	&	\, 0.0	&	22.9	&	77.1	\\
6.0	&	10.0\,\,\,	&	\, 0.0	&	11.4	&	88.6	\\
10.0\,\,\,	 &	70.0\,\,\,&	\, 0.0	&	22.6	&	77.4	\\
\cutinhead{$S_{250}/S_{24}$}
1.0	&	3.0	&	\, 0.0	&	33.1	&	66.9	\\
3.0	&	7.0	&	\, 4.6	&	\, 1.9	&	93.5	\\
7.0	&	10.0\,\,\,	&	\, 5.4	&	11.4	&	83.2	\\
10.0\,\,\,	&	20.0\,\,\,	&	\, 6.4	&	37.0	&	56.6	\\
20.0\,\,\,	&	30.0\,\,\,	&	45.2	&	34.7	&	20.1	\\
30.0\,\,\,	&	40.0\,\,\,	&	61.6	&	36.3	&	\, 2.1	\\
40.0\,\,\,	&	50.0\,\,\,	&	58.6	&	41.4	&	\, 0.0	\\
50.0\,\,\,	&	60.0\,\,\,	&	65.7	&	26.3	&	\, 8.0	\\
60.0\,\,\,	&	90.0\,\,\,	&	62.5	&	37.5	&	\, 0.0	\\
90.0\,\,\,	&	200.0\quad\quad	&	50.9	&	48.5	&	\, 0.6
\enddata
\tablenotetext{}{We list the percentages of SFGs, Composites, and AGN in each color bin shown in Figure \ref{fig:col_dist}.}
\end{deluxetable}

\section{Summary}
\label{sec:seven}
We have decomposed mid-IR spectroscopy to robustly determine the strength of an AGN, classified as the fraction of mid-IR luminosity due to power-law continuum emission, in a sample of 343 high redshift (U)LIRGs. 
We define three general classifications: SFGs (\fAGN$<$0.2), composites (\fAGN=0.2-0.8), and AGN (\fAGN$>$0.8).
Based on these mid-IR classifications, 
we have created three publicly available template libraries designed for use with high redshift LIRGs and ULIRGs. The appropriate library depends on the data available to the user:
\begin{enumerate}
\item MIR-based Library. This is ideal if information about the mid-IR power source is available, but little far-IR data is available. 
\item Color-based Library. These are ideal for high redshift sources which only have photometric data available.
\item Comprehensive Library. This library is based on comprehensive intrinsic galaxy information. We have used it to study dust emission trends with \fAGN, redshift, and $L_{\rm IR}$. Choosing the appropriate template from this library requires knowledge about a source's $L_{\rm IR}$ and redshift.
\end{enumerate}
Using our empirical templates, we find
\begin{itemize}
\item SFGs are remarkably similar from $z\sim0.3-2.8$. The shape of the mid-IR and far-IR emission is nearly identical for the three SFG templates from the Comprehensive Library, and the dust temperatures ($T_c$, $T_w$) and normalizations ($L_{\rm cold}/L_{\rm IR}$) are consistent. Furthermore, the colors of these templates are similar to colors of low redshift LIRGs from GOALS, indicating that local analogs exist for high redshift star forming LIRGs and ULIRGs, albeit at a slightly lower $L_{\rm IR}$. A detailed comparison of the dust emission of high redshift (U)LIRGs and their local analogs will be discussed in an upcoming study (Kirkpatrick et al.\,2016, in preparation).
\item For composites and AGN, the cold dust temperature, $T_c$, changes with $L_{\rm IR}$ and redshift, but it is not affected by the strength of the AGN as $T_c$ arises from the host galaxy.
\item The warm dust temperature, $T_w$, and the relative amount of cold dust emission, $L_{\rm cold}/L_{\rm IR}$ are a strong function of \fAGN. As the AGN grows more luminous, it heats more of the dust to warmer temperatures, eventually outshining the cold dust component. \fAGN=0.6 is an interesting threshold where $T_{w}$ peaks and $L_{\rm cold}/L_{\rm IR}$ begins to decline.
\item \fAGN is related to the total amount of $L_{\rm IR}$ from AGN heating, \fTOT, by a 2nd-degree polynomial. Due to the quadratic nature of the relationship, an AGN does not significantly contribution to $L_{\rm IR}$ until \fAGN$>$0.6. \fTOT is useful to correct the amount of $L_{\rm IR}$ attributable to star formation and obtain more accurate star formation rates. 
\item In general, we find that composites are a true mix of SFGs and AGN, and may represent a transition between the two. A merger or other instability triggers the growth of an AGN which can heat the dust below $\sim$40\,$\mu$m and suppress PAH emission, producing higher \fAGN values. However, the AGN does not manifest itself on the far-IR emission until \fAGN$>$0.6, and eventually, the AGN-heated dust outshines the diffuse dust.
\item We estimate how prevalent AGN and composites are at different 24\,$\mu$m selection thresholds, and find that $>$40\% of a sample selected at $S_{\rm 24} > 0.1\,$mJy may be hosting a buried AGN. Composites and AGN have at least $>$20\% of $L_{\rm IR}$ due to AGN heating, illustrating the necessity of accounting for AGN heating when studying dust emission or IR-based SFRs at high redshift.
\end{itemize}

Our infrared analysis will be applicable for forthcoming data from JWST. MIRI will provide medium resolution spectroscopy from 5-28\,$\mu$m. Our spectral decomposition technique requires coverage of the PAH complexes from 6-13\,$\mu$m. With the coverage of MIRI, our mid-IR decomposition technique can be used to 
identify mid-IR AGN out to $z\sim2$. We have also demonstrated that the color $S_8/S_{3.6}$, obtainable with JWST, can be used to separate AGN from SFGs in the range $z=0-2.8$. Future observations with JWST can reach deeper 24\,$\mu$m limits to determine the prevalence of AGN and composites in these samples.

\acknowledgements
We thank the anonymous referee for his/her thoughtful comments which have helped to improve the clarity and impact of this work.
We acknowledge support from NSF AAG grants AST-1312418 and AST-1313206. S. Stierwalt also gratefully acknowledges support from the L'Oreal USA For Women in Science program. 
T. D\'{i}az-Santos. acknowledges support from ALMA-CONICYT project 31130005 and FONDECYT project 1151239.

\renewcommand{\thefigure}{A-\arabic{figure}}
\setcounter{figure}{0}

\appendix

\section{A. Alternate Fitting Methods}
\label{app:alternate}
\begin{figure}
\centering
\includegraphics[width=2.3in]{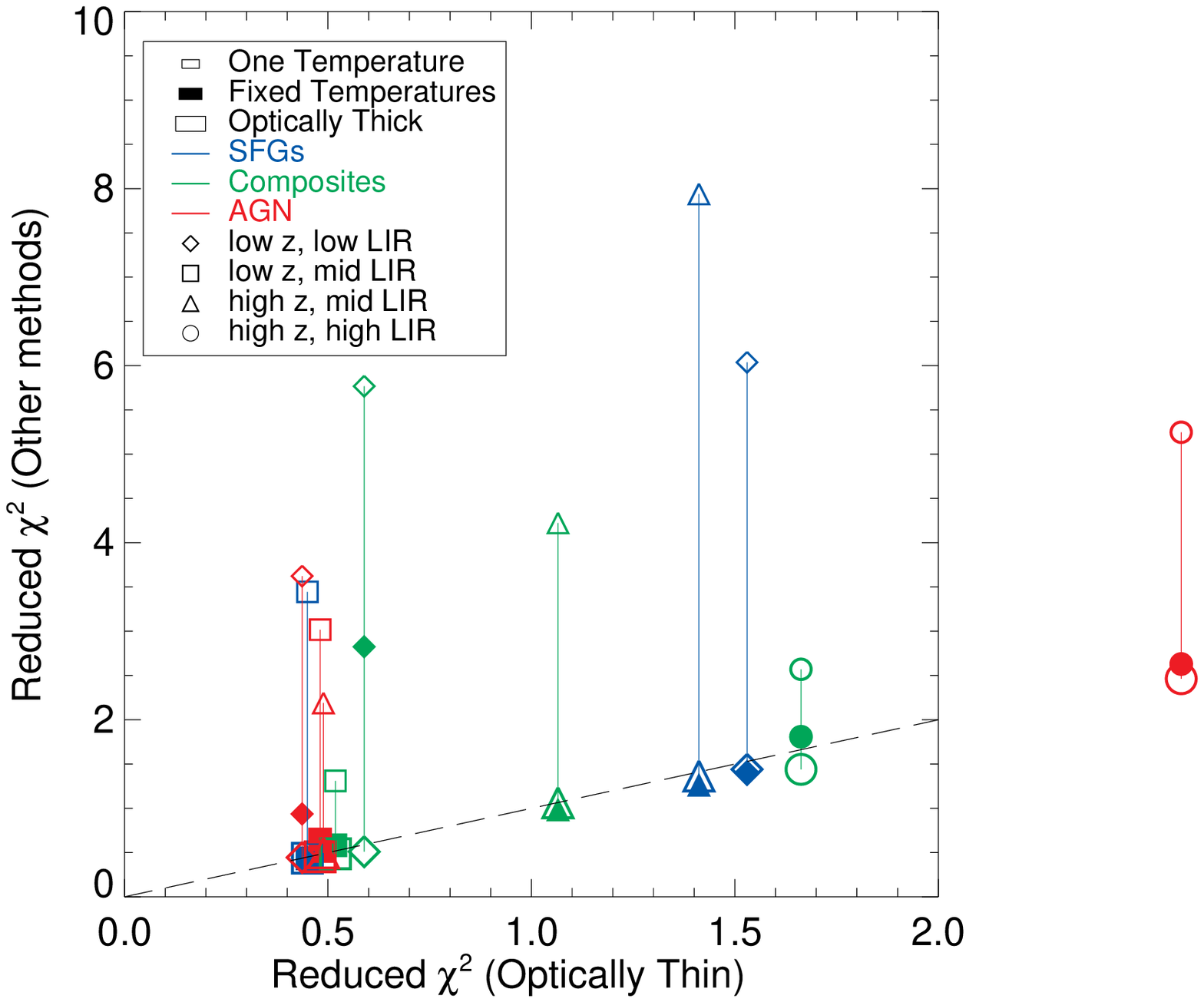}
\includegraphics[width=2.3in]{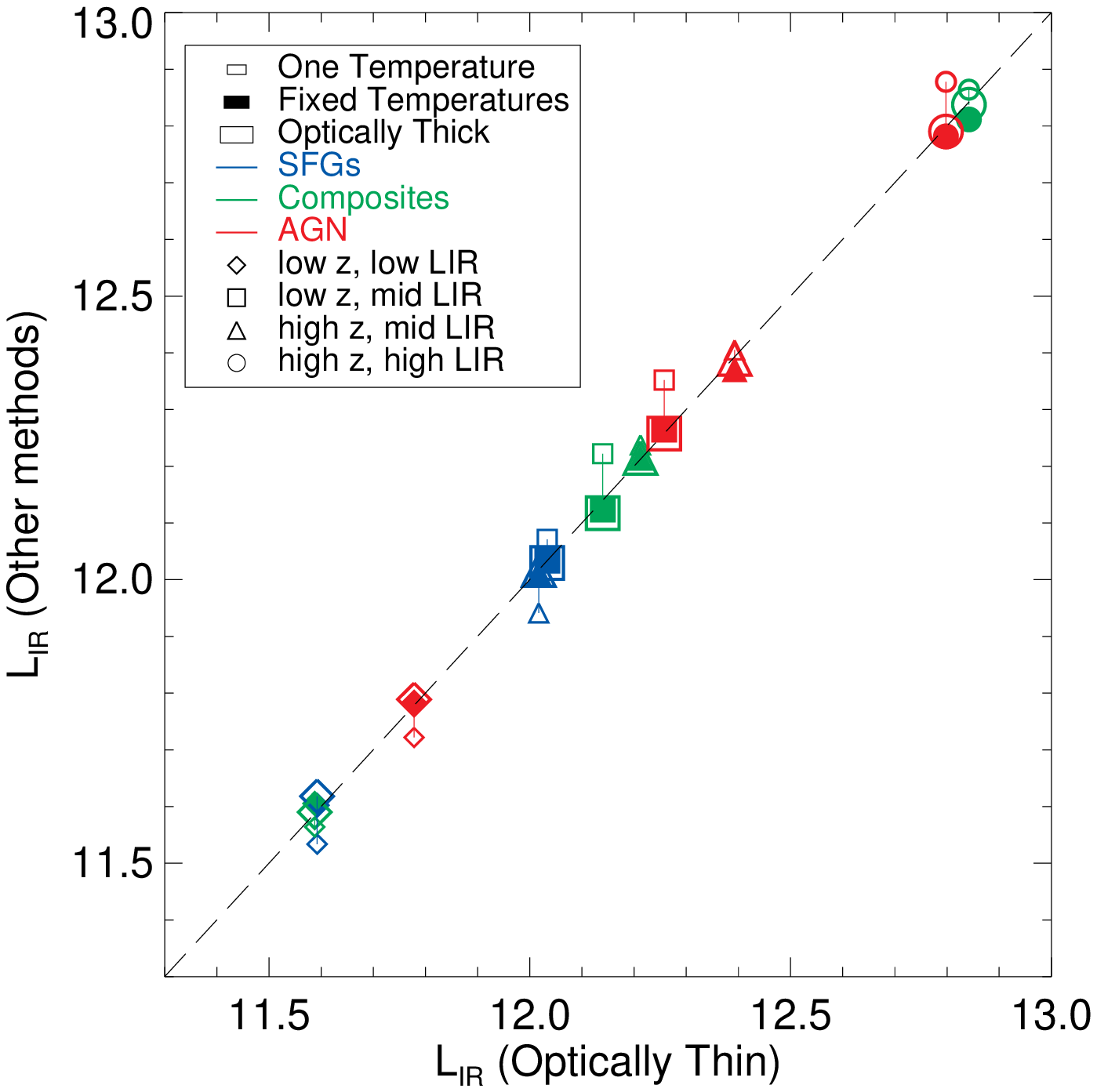}
\includegraphics[width=2.3in]{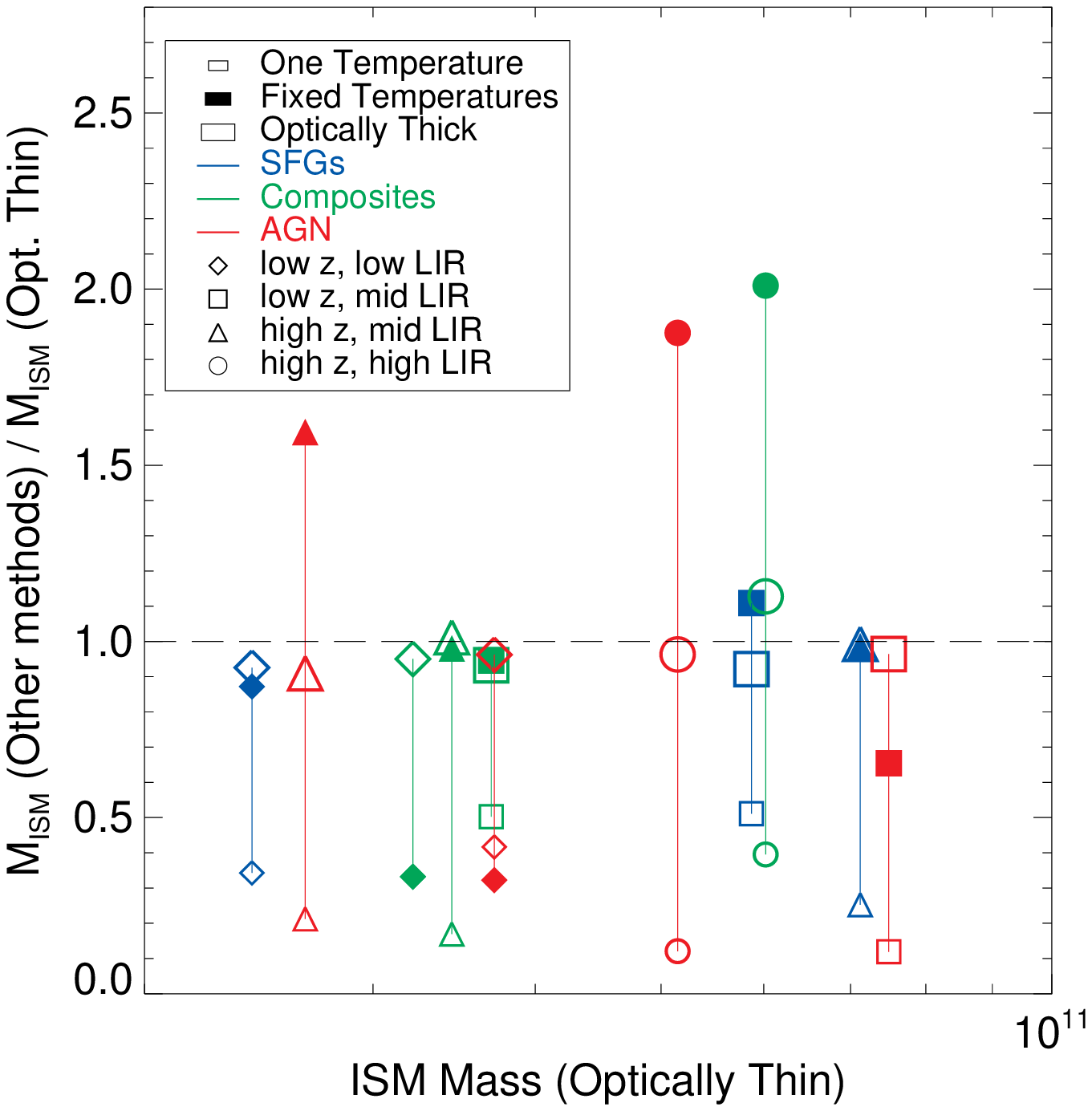}
\caption{{\it Left Panel--} Comparison of the reduced $\chi^2$ values from our different fitting methods. Colors and symbols correspond to each template from the Comprehensive Library. Filled symbols are the reduced $\chi^2$ values derived when the two temperature components are held fixed; large open symbols are derived using the optically thick assumption; small open symbols are derived with only a one temperature MBB, instead of two temperatures. We overplot a one-to-one relation as the dashed line. The 1T MBB method produces the worst reduced $\chi^2$ values, while there is smaller difference between the reduced $\chi^2$ values using the optically thick or optically thin dust assumption.
{\it Middel Panel--} We compare the $L_{\rm IR}$ values calculated from each method. $L_{\rm IR}$ is essentially independent of the particular far-IR fitting method.
{\it Right Panel--} We compare the ISM masses derived from each fitting method. The optically thick and optically thin assumptions produce consistent ISM masses, while the one temperature fitting method results in significantly lower ISM masses. \label{fig:chisq}}
\end{figure}

We fit the far-IR SED with an optically thin 2T MBB to construct our templates, and we now discuss whether this fitting method is optimal for determining dust temperatures and $L_{\rm IR}$. We use the subsamples in the Comprehensive Library to explore three alternate fitting methods:
\begin{enumerate}
\item Optically thick dust
\item Fixed dust temperatures
\item One temperature MBB
\end{enumerate}
For each method, we follow the same fitting procedure outlined in Section \ref{sec:four}, and we compare the results with the optically thin 2T MBB fits used to create our templates.
We quantify the goodness of the fits with the reduced $\chi^2$ statistic, and we compare the reduced $\chi^2$ values in the left panel of Figure \ref{fig:chisq}.

$L_{\rm IR}$ is typically a desired quantity when fitting far-IR data. We compare $L_{\rm IR}$ calculated from each of the three alternate fitting methods with our template $L_{\rm IR}$s (Table \ref{tbl:temp_stats}) in the middle panel of Figure \ref{fig:chisq}. We find no significant difference for any of the templates, showing that $L_{\rm IR}$ is robust against these particular fitting methods. However, $L_{\rm IR}$ is not the only useful parameter that can be derived from fitting far-IR photometry with a model; another commonly calculated quantity is ISM mass.
We demonstrate how a particular far-IR fitting technique affects the derived ISM mass in the right panel of Figure \ref{fig:chisq}. For each template, we calculate the ISM mass at 850\,$\mu$m, which is in the Rayleigh-Jeans tail of the dust emission and is a more reliable tracer of the ISM mass \citep{scoville2014}. We use the following equation:
\begin{equation}
M_{\rm ISM}=\frac{\lambda^2 L_\nu}{8\pi k\,\kappa_{\rm ISM}\,T_c}
\end{equation}
$\kappa_{\rm ISM}$ is the dust opacity per grain and is related to the opacity $\tau_\nu$. $\tau_{250}/N_{\rm H}$ has been recently measured by the Planck Collaboration \citep{planck2011}, and from that value, \citet{scoville2014} calculate $\kappa_{\rm ISM}(\nu_{250})$:
\begin{equation}
\kappa_{\rm ISM}(\nu_{250})=\frac{\tau_{250}}{N_{\rm H}1.36\,{\rm m}_{\rm H}}
\end{equation}
where $N_{\rm H}$ is the column density of hydrogen.
$\kappa_{\rm ISM}(\nu_{250})$ can then be scaled to 850\,$\mu$m:
\begin{equation}
\kappa_{\rm ISM}(\nu_{850})=\kappa_{\rm ISM}(\nu_{250})\times\left(\frac{250}{850}\right)^{-\beta}
\end{equation}

\subsection{A.1 Optically Thick Dust}
The optically thin dust approximation is commonly adopted with a limited number of data points, but it might not be an accurate assumption at $\lambda<<100\,\mu$m, particularly in starbursts. We test what effect using the full optically thick equation has on the dust temperatures by fitting to the far-IR data points of each template in the Comprehensive Library. We fit
\begin{equation}
S_\nu=a_1\times (1-e^{-\tau(\nu)}) \times B_\nu(T_{\rm warm})+a_2\times(1-e^{-\tau(\nu)}) \times B_\nu(T_{\rm cold})
\end{equation}
where $\tau(\nu)=(\nu/\nu_0)^\beta$. We use $\beta=1.5$, and we assume $\nu_0=300\,$GHz ($\lambda_0=100\,\mu$m). The optically thick equation produces reduced $\chi^2$ values consistent with the optically thin fitting. As for the physical parameters, we find that the optically thick equation has a negligible effect on $T_c$ and $L_{\rm cold}/L_{\rm IR}$, since the cold dust is presumably optically thin, but increases the derived $T_w$ by $\sim20\,$K. However, the exact value of $T_w$ ultimately has little effect on $L_{\rm IR}$, and the warm dust component accounts for 
$\sim1\%$ of the ISM mass. 
Since $\chi^2$, $L_{\rm IR}$, and $M_{\rm ISM}$ do not change significantly when assuming optically thin dust, we recommend using the optically thin approximation for simplicity.

\subsection{A.2 Fixed Dust Temperature}
We experiment with holding the dust temperatures fixed which is another useful technique when limited data are available. We hold the temperatures fixed to the average $T_c$ and $T_w$ values for the SFGs, Composites, and AGN, separately. In general, holding the temperatures fixed has little effect on the relative normalizations of the dust component, so the ratio $L_{\rm cold}/L_{\rm IR}$ is approximately constant compared with when the dust temperatures are allowed to vary. $M_{\rm ISM}$ is significantly higher for the 
Composite4, AGN3, and AGN4 templates, and lower for the AGN1 and AGN2 templates. The AGN templates show the most increase of $T_c$ with $L_{\rm IR}$ and redshift, and this is not captured by holding the temperatures fixed, producing incongruous ISM masses. 

We also attempt to fit a 3T MBB, as this may be more physically appropriate, particularly for the AGN sources. In this case, the coldest dust component comes from the diffuse ISM, a warmer component is due to heating from star forming regions, and a hot component is due to heating by an AGN. In order to achieve good fits, we had to assume dust temperatures. Based on the temperatures of the diffuse component and star forming regions in the local Universe, we assumed $T_c=20\,$K, $T_w=40\,$K, and $T_h=100\,$K \citep[e.g.,][]{clemens2013}. The results produced good reduced $\chi^2$ fits and consistent $L_{\rm IR}$s, but we do not advocate this technique as it requires assumptions about the dust temperatures which may not hold at high redshift.

\subsection{A.3 One Temperature}
Finally, we test how good of a fit we can achieve with only a 1T MBB, which is commonly adopted in the literature due to incomplete photometric coverage of the far-IR. In this case, the reduced $\chi^2$ values are typically poor ($>2$). This result occurs because we are fitting the wavelength range $\lambda \sim 20 - 300\,\mu$m, and a 1T MBB will necessarily be biased to warmer dust temperatures by including this much data.  The 1T
MBB produces consistently lower ISM masses, typically 60-70\% lower than optically thin 2T MBB method, due to both the difference in $T_c$ and the extrapolated $L_{850}$. If only SPIRE data is available, we recommend adding in a warm dust component with a fixed temperature in order to ensure the cold dust temperature is not biased to warmer wavelengths \citep[e.g.,][]{kirkpatrick2014a}. A 2T MBB, even with a fixed warm dust component, is optimal for fitting the peak of the SED and determining $T_c$.

\hfill\break

\newpage
\section{B. Template Libraries}
\label{app:temp}
We present the complete data sets that comprise each template. The spectra are plotted as lines and the photometry as open circles. We use a different color for each source in a given subsample. We plot any available submillimeter data as the open squares. These data were not included in the fit, since they are not available for all sources, but the are plotted to illustrate how well the Rayleigh-Jeans tails of our templates agree with observations. The templates and associated uncertainties are plotted as the thick black lines and grey shaded regions. We also plot the warm modified blackbody and cold modified blackbody from Equation \ref{eq:BB} as the long dashed and dotted lines, respectively. We remind the reader that the three libraries are not independent as they all contain the same sources divided according to different criteria. {\color{red} Due to arXiv size limitations, full libraries cannot be shown here. They will appear in the published ApJ paper.}

\end{document}